# Optimal Biological Dose Combination Finding: A Design Roadmap and Robust Cross-Indication Bayesian Borrowing

Ayon Mukherjee[1,*] Kentaro Takeda[2] James M.S. Wason[1]

[1] Population Health Sciences Institute, Newcastle University, Newcastle upon Tyne, UK
[2] Quantitative Science and Evidence Generation, Astellas Pharma Global Development Inc., USA

[*] Correspondence: Ayon Mukherjee, Population Health Sciences Institute, Newcastle University, Newcastle upon Tyne, UK
(ayon.mukherjee@newcastle.ac.uk)

**Abstract**

Early-phase combination oncology trials increasingly seek the optimal biological dose combination (OBDC) rather than a maximum tolerated dose combination, in line with the FDA's Project Optimus initiative. Existing Bayesian OBDC designs span rule-based, model-assisted, and model-based paradigms, but no unified framework exists for choosing among them, and none supports borrowing information across indications sharing a combination regimen. We propose a corrected three-by-two taxonomy of OBDC designs by mechanism and objective, a rule-based design (Ji3+3-Comb) closing a documented gap in transparent, model-free combination dose-finding, and a Bayesian Hierarchical Utility-based Cross-indication (BHUC) design that borrows information across indications via a robust mixture prior while discounting non-exchangeable information. We derive the shared utility function as the Bayes-optimal decision under a linear clinical loss, and prove that BHUC's mixture-prior posterior weight automatically vanishes as two indications' true rates become discordant, formalizing its robustness to non-exchangeable borrowing, and complement this asymptotic guarantee with an exact, sample-size-free ceiling on borrowed influence that holds even before any own-indication data accrue. A roadmap links trial features to design choice, and a 5000-replication simulation study, sensitivity analyses, and a case study built from a published phase Ib trial show that model-assisted designs offer the most robust safety–efficacy trade-off, while BHUC improves correct selection over independent per-indication designs even under discordance. We conclude with concrete, protocol-actionable recommendations for pharmaceutical and trial biostatisticians, and directions for future methodological and computational development.

## 1 Introduction

Combination regimens—pairing two therapeutic agents to exploit synergy or distribute toxicity across mechanisms—are foundational to modern oncology drug development (Humphrey et al. 2011). For cytotoxic chemotherapy, early-phase trials have traditionally sought a maximum tolerated dose combination (MTDC), implicitly assuming that efficacy increases monotonically with dose. Targeted agents and immunotherapies routinely violate this assumption: efficacy frequently plateaus, or even declines, well below the maximum tolerable exposure (Thall and Cook 2004; Hamberg and Verweij 2009), and toxicity may be delayed, immune-mediated, or driven by a different pharmacological mechanism than the anti-tumour effect. The U.S. Food and Drug Administration's Project Optimus initiative and its accompanying dose-optimization guidance formalize the resulting shift in objective: from the MTDC to the *optimal biological dose combination* (OBDC), the most conservative dose pair that is simultaneously safe and sufficiently active (U.S. Food and Drug Administration 2024; Shah et al. 2021; Zirkelbach et al. 2022). The FDA's parallel draft guidance on Bayesian methodology further encourages model-based and adaptive designs that make full use of accruing toxicity and efficacy data (U.S. Food and Drug Administration 2026), and the ICH M15 guideline on model-informed drug development (International Council for Harmonisation of Technical Requirements for Pharmaceuticals for Human Use (ICH) 2026) and the estimands framework of ICH E9(R1) (International Council for Harmonisation of Technical Requirements for Pharmaceuticals for Human Use (ICH) 2019; Mukherjee et al. 2025) both call for pre-specified, reproducible decision rules linking trial data to dosing recommendations.

A substantial methodological literature now targets the OBDC directly for two-agent combinations. Model-based designs, exemplified by the Bayesian efficacy–toxicity trade-off design EffTox (Thall and Cook 2004) and its successful clinical implementation in the MATCHPOINT trial (Brock et al. 2017), fit explicit joint probability surfaces for toxicity and efficacy across the dose grid and update them via full posterior inference. Rule-based (model-free) designs such as the up-and-down and interval-based approaches for single agents (Babb et al. 1998; Ivanova and Wang 2004; Wang and Ivanova 2005; Fan et al. 2009) and the Joint i3+3 design for cellular therapies (Lin and Ji 2020) instead tabulate simple, auditable escalation rules directly in terms of observed toxicity and efficacy rates. Between these extremes, model-assisted designs pre-compute closed-form Bayesian decision boundaries—in the spirit of the Bayesian Optimal Interval (BOIN) design (Liu and Yuan 2015) and its combination extension (Lin and Yin 2017) and utility-based single-agent extension BOIN12 (Lin et al. 2020)—and pair them with a lightweight posterior utility ranking, so that the entire decision logic can be tabulated before the trial begins. Comb-BOIN12 (Lu et al. 2025) is the current state of the art in this class for two-agent OBDC finding. A separate line of work targets model selection uncertainty directly: the BODC design of Wang et al. (2024) performs Bayesian model averaging across candidate dose–efficacy curves, while COMIC (Chen et al. 2025) and uTPI-Comb (Liang et al. 2024) use zone- and candidate-set-based utility comparisons to integrate toxicity and efficacy without full likelihood-based model fitting. Complementary designs incorporate patient-level covariates or delayed outcomes via flexible machine-learning components (Zhao et al. 2024; Chung et al. 2025; Chen et al. 2024) or continuous dose curves (Tighiouart et al. 2014, 2017; Jimenez et al. 2019, 2020; Jimenez and Tighiouart 2022; Jimenez and Zheng 2023), and classical two-dimensional dose-finding methods (Mander and Sweeting 2015; Cai et al. 2014; Guo and Yuan 2017; Yin and Yuan 2009; Pan et al. 2020; Wages et al. 2011; Le Tourneau et al. 2009; Neuenschwander et al. 2008) continue to inform prior specification and boundary calibration for these newer designs.

Despite this progress, three gaps limit the practical uptake of Bayesian OBDC designs. We identified these gaps from the two-agent OBDC design literature discussed throughout this section and its citing and cited works, located through targeted searches of PubMed and Google Scholar (search terms including "optimal biological dose combination", "Bayesian dose optimization combination", and "two-agent phase I/II design", through September 2026); this is a targeted rather than a systematic search, so the gaps below are stated as being, to our knowledge, unaddressed in this literature rather than as exhaustively verified absences. First, existing classifications of dose-finding designs conflate the *mechanism* by which a design reaches a decision (rule-based, model-assisted, or model-based) with the *objective* it targets (an MTDC versus a utility-integrated

OBDC), producing overlapping, non-exhaustive categories that make it difficult for a trial team to map their scenario onto an appropriate design. Second, to our knowledge no rule-based, fully model-free design has yet been proposed for two-agent OBDC estimation, despite clinicians' well-documented preference for simple, transparent decision rules in early-phase trials (Le Tourneau et al. 2009): the model-free paradigm has been developed extensively for single-agent MTD-finding but not carried through to the utility-integrated, two-agent OBDC setting. Third, as combination regimens are increasingly developed simultaneously across several tumour types or histologies, to our knowledge no existing OBDC design formally borrows information across indications while protecting against the harm that borrowing can cause when the indications are not exchangeable (Schmidli et al. 2014).

This article addresses all three gaps and is written with pharmaceutical and trial biostatisticians in mind: our aim throughout is a set of designs, proofs, and operating-characteristic evidence that a statistician planning a two-agent combination trial can act on directly rather than merely read about. Section 2 gives a corrected, mutually exclusive taxonomy of Bayesian OBDC designs by mechanism and objective. Section 4 introduces two new designs: Ji3+3-Comb, a rule-based two-agent OBDC design that closes the first gap, and the Bayesian Hierarchical Utility-based Cross-indication (BHUC) design, a model-assisted design with a robust mixture-prior hierarchical borrowing mechanism that closes the third gap; both are benchmarked against Comb-BOIN12 (Lu et al. 2025) and a fast Laplace-approximation implementation of a joint model-based design in the spirit of EffTox. Both new designs are given a formal theoretical grounding: Section 3 derives the shared utility function as the Bayes-optimal decision under a linear clinical loss (Proposition 3.1), and Section 4.3 proves that BHUC's mixture-prior borrowing mechanism automatically discounts information from a discordant indication in the large-sample limit (Proposition 4.1) and, complementing this asymptotic result, an exact finite-sample ceiling on the pooled component's influence that holds at every sample size, including at an untested cell (Proposition 4.2). Section 5 distils these results into a practical roadmap for design selection. Sections 6 and 7 report a 5000-replication simulation study and accompanying prior- and model-assumption sensitivity analyses. Section 8 illustrates all designs on a grid constructed from the published dose-escalation results of a real two-agent phase Ib oncology trial. Section 9 discusses practical recommendations, regulatory context, limitations, and future methodological directions, including how the hierarchical borrowing mechanism proposed here could be extended to a fully exchangeable random-effects model as more indications and historical trials accumulate, and how artificial-intelligence and machine-learning methods might further improve the efficiency with which the OBDC itself is identified.

## 2 A Taxonomy of Bayesian Designs for Optimal Biological Dose Combination Finding

We classify two-agent OBDC designs along two orthogonal axes. The *mechanism* axis describes how a design converts accruing data into a dosing decision: a *rule-based* (model-free) design compares observed toxicity and efficacy rates directly against pre-tabulated boundaries with no probability model beyond the immediate binomial likelihood at the current cell; a *model-assisted* design uses a simple conjugate probability model (typically an independent, or quasi-independent, Beta–Binomial model at each cell) to compute closed-form Bayesian decision boundaries and posterior tail probabilities that can be tabulated before the trial begins; a *model-based* design fits a parametric regression surface (logistic, Emax, or copula) jointly across the whole dose grid, borrowing strength between cells through the regression structure and requiring posterior computation (Markov chain Monte Carlo or an approximation to it) at each interim decision. The *objective* axis distinguishes designs that target only a maximum tolerated combination from those that integrate toxicity and efficacy into a single utility function to target the OBDC. Table 1 and Figure 1 place the designs discussed in this article, together with the two we propose, in this $3 \times 2$ scheme; unlike previous informal groupings, the two axes are defined independently, so every cell is well defined and no design can occupy more than one cell.

**Table 1** A $3 \times 2$ taxonomy of Bayesian dose-finding designs for two-agent combinations, classified by decision mechanism and by objective. Designs introduced in this article are marked with $^*$.

| Mechanism | Toxicity-only objective (MTDC) | Utility-integrated objective (OBDC) |
|---|---|---|
| Rule-based (model-free) | Up-and-down / 3+3-type combination rules (Ivanova and Wang 2004; Wang and Ivanova 2005; Fan et al. 2009) | **Ji3+3-Comb**$^*$ (§4.1); single-agent Joint i3+3 (Lin and Ji 2020) |
| Model-assisted | Comb-BOIN (Lin and Yin 2017); Keyboard-combo (Pan et al. 2020) | Comb-BOIN12 (Lu et al. 2025); **BHUC**$^*$ (§4.3); uTPI-Comb (Liang et al. 2024); COMIC (Chen et al. 2025) |
| Model-based | Copula / continuation-ratio combination CRM (Yin and Yuan 2009; Wages et al. 2011) | EffTox (Thall and Cook 2004; Brock et al. 2017); BODC (Wang et al. 2024); EffTox-approx (§4.4) |

|  | Toxicity-only (MTDC) | Utility-integrated (OBDC) |
|---|---|---|
| Rule-based | 3+3 / up-and-down combination rules | **Ji3+3-Comb** (proposed) |
| Model-assisted | Comb-BOIN Keyboard-combo | Comb-BOIN12 **BHUC (proposed)** uTPI-Comb; COMIC |
| Model-based | Copula / CRM-combo | EffTox / EffTox-approx BODC |

**Figure 1** Visual layout of the taxonomy in Table 1. Shaded cells (right column) target the optimal biological dose combination directly through an integrated utility function; designs proposed in this article are set in bold.

## 3 Notation, Utility Function, and Admissibility

Let two agents be administered at dose levels $a = 1, \ldots, J$ and $b = 1, \ldots, K$, defining a $J \times K$ grid of candidate combinations $(a, b)$. For a cohort of patients treated at $(a, b)$, let $n_{ab}$ denote the number of patients treated, $y^T_{ab}$ the number experiencing a dose-limiting toxicity, and $y^E_{ab}$ the number experiencing an efficacy response, with true underlying probabilities $p^T_{ab}$ and $p^E_{ab}$. Throughout the simulation study and case study of Sections 6–8, each simulated patient's toxicity and efficacy outcomes are generated as conditionally independent Bernoulli variables given the assigned dose combination and its true $(p^T_{ab}, p^E_{ab})$; this is a simplifying assumption made for transparency and computational speed at the scale of a 5000-replication, multi-scenario study, and it fully determines the joint outcome distribution used throughout (rather than leaving it unspecified), but it does not allow for within-patient association between toxicity and efficacy beyond what the two marginal rates induce, and relaxing it, for example via a copula linking the two margins, is a natural extension we return to in Section 9. Let $\phi_T$ and $\phi_E$ denote the maximum acceptable toxicity rate and the minimum acceptable efficacy rate, respectively. A cell is *admissible* if it satisfies both a safety and an efficacy posterior-probability constraint,

$$\begin{aligned} \Pr(p_{ab}^T > \phi_T \mid \mathcal{D}_{ab}) \le C_T, \\ \Pr(p_{ab}^E < \phi_E \mid \mathcal{D}_{ab}) \le C_E, \end{aligned} \tag{1}$$

where $\mathcal{D}_{ab}$ is the data accrued at cell $(a, b)$ and $C_T, C_E$ are pre-specified cut-offs (we use $C_T = C_E = 0.90$ throughout). Among admissible cells, the utility function

$$u(a, b) = p_{ab}^E - w_T \, p_{ab}^T \tag{2}$$

trades off efficacy against toxicity through a pre-specified weight $w_T > 0$ (we use $w_T = 0.5$ as a base case and examine its influence in Section 7); the OBDC is the admissible cell maximizing $u(a, b)$.

Equation (2) is not an ad hoc scoring rule: it is, up to its single free parameter $w_T$, the essentially unique per-patient loss structure under which the OBDC is a Bayes-optimal decision, subject to the admissibility constraints of Equation (1).

**Proposition 3.1** (Utility as Bayes risk under linear clinical loss)**.** *Let $(T, E) \in \{0, 1\}^2$ denote a patient's toxicity and efficacy indicators at dose $(a, b)$, with $T \perp\!\!\!\perp E \mid (a, b)$, $\Pr(T = 1 \mid a, b) = p_{ab}^T$ and $\Pr(E = 1 \mid a, b) = p_{ab}^E$ as in Section 3. Suppose the clinical loss of treating a patient at $(a, b)$ is any additively separable function of the two binary outcomes, $L(T, E) = \ell_T(T) - \ell_E(E)$, with $\ell_T, \ell_E$ non-decreasing on $\{0, 1\}$ and normalized so that $\ell_T(0) = \ell_E(0) = 0$ and $\ell_E(1) = 1$ (fixing the scale on which loss is measured) and $\ell_T(1) = w_T$ (the relative cost of a toxicity event on this scale). Then* (i) *the posterior expected loss of treating a patient at $(a, b)$ is $\mathbb{E}[L(T, E) \mid \mathcal{D}_{ab}] = w_T \, \mathbb{E}[p_{ab}^T \mid \mathcal{D}_{ab}] - \mathbb{E}[p_{ab}^E \mid \mathcal{D}_{ab}] = -\mathbb{E}[u(a, b) \mid \mathcal{D}_{ab}]$, where $u(a, b)$ is exactly the utility of Equation (2); and* (ii) *the admissible cell minimizing posterior expected loss is exactly the admissible cell maximizing posterior mean utility, so the OBDC of Equation (2) is the unique Bayes-optimal action, among admissible cells, for every loss function of this additively separable, binary-outcome form with toxicity cost $w_T$.*

*Proof.* Any non-decreasing function on $\{0, 1\}$ with value 0 at 0 is determined by its value at 1, so $\ell_T(T) = w_T T$ and $\ell_E(E) = E$ pointwise, giving $L(T, E) = w_T T - E$. Taking expectations and using $T \perp\!\!\!\perp E \mid (a, b)$,

$$\mathbb{E}[L(T, E) \mid \mathcal{D}_{ab}] = w_T \, \mathbb{E}[p_{ab}^T \mid \mathcal{D}_{ab}] - \mathbb{E}[p_{ab}^E \mid \mathcal{D}_{ab}],$$

which is $-u(a, b)$ evaluated at the posterior means, or $-\mathbb{E}[u(a, b) \mid \mathcal{D}_{ab}]$ under posterior expectation of the (linear) function $u$. Minimizing this over the constraint set of admissible cells defined by Equation (1) is therefore equivalent to maximizing posterior mean utility over the same set, which is exactly the OBDC selection rule of Section 3. The admissibility constraints of Equation (1) are imposed separately from, rather than traded off within, this loss, formalizing the clinical requirement that an unacceptably toxic or inadequately active dose cannot be selected merely because it scores well on the other attribute. □

Proposition 3.1 clarifies the role of $w_T$: rather than an arbitrary tuning constant, it is the only free parameter distinguishing one member from another within the entire family of loss functions that are additively separable in the two binary trial outcomes, so specifying $w_T$ is both necessary and sufficient to determine the Bayes-optimal design objective within this natural family.

For the model-assisted designs we follow the quasi-binomial construction of Comb-BOIN12 (Lu et al. 2025): writing $x_{ab} = \{y_{ab}^E + w_T(n_{ab} - y_{ab}^T)\}/(1 + w_T)$ as a continuised "success" count, a $\mathrm{Beta}(a_0, b_0)$ prior gives the conjugate posterior $x_{ab} \mid n_{ab} \sim \mathrm{Beta}(a_0 + x_{ab},\, b_0 + n_{ab} - x_{ab})$ for the standardized utility $u_{ab}^* = \{u(a,b) + w_T\}/(1 + w_T) \in [0,1]$, so that both the terminal posterior mean utility and the in-trial ranking statistic $\Pr(u_{ab}^* > u_b^* \mid \mathcal{D}_{ab})$ against the pre-specified benchmark $u_b^* = \{(\phi_E - w_T\phi_T) + w_T\}/(1 + w_T)$ have closed form. Escalation and de-escalation boundaries $\lambda_e$ and $\lambda_d$ for the observed toxicity rate $\hat{p}_{ab}^T = y_{ab}^T/n_{ab}$ are obtained from the standard BOIN optimality criterion (Liu and Yuan 2015) with target $\phi_T = 0.35$ and equivalence interval $(0.6\phi_T,\, 1.4\phi_T)$, giving $\lambda_e = 0.2763$ and $\lambda_d = 0.4189$ throughout this article.

## 4 Proposed and Benchmark Designs

### 4.1 Ji3+3-Comb: a rule-based two-agent OBDC design

Ji3+3-Comb extends the single-agent Joint i3+3 design (Lin and Ji 2020) to two-agent combinations, closing the gap identified in Section 1. At the current cell $(a, b)$, with equivalence margin $\varepsilon = 0.05$, the observed toxicity rate is classified as *overdosing* if $\hat{p}_{ab}^T > \phi_T + \varepsilon$, *underdosing* if $\hat{p}_{ab}^T < \phi_T - \varepsilon$, and *proper* otherwise; the observed efficacy rate is classified as *low* if $\hat{p}_{ab}^E < \phi_E - \varepsilon$ and *adequate* otherwise. If toxicity is classified as overdosing, the design de-escalates to whichever of the two lower neighbouring cells $(a-1, b)$, $(a, b-1)$ has the lower observed toxicity rate; otherwise, if efficacy is low, the design escalates to whichever of $(a+1, b)$, $(a, b+1)$ has the lower observed toxicity rate (agent-specific toxicity being used here only as a tie-breaker, since no efficacy data yet exist at the untested neighbour); otherwise the design stays at $(a, b)$. At the maximum sample size, the OBDC is selected as the visited, tested cell with acceptable observed toxicity and adequate observed efficacy that maximizes the empirical utility $\hat{u}(a, b) = \hat{p}_{ab}^E - w_T\hat{p}_{ab}^T$. Because every decision is a direct comparison of observed proportions against fixed boundaries, the entire rule can be tabulated in a trial protocol without any posterior computation, matching the transparency of the classical 3+3 design while explicitly targeting the OBDC rather than the MTDC.

### 4.2 Comb-BOIN12: model-assisted benchmark

We use Comb-BOIN12 (Lu et al. 2025) as the model-assisted benchmark representing current practice. At each interim decision, the direction of movement is determined by comparing $\hat{p}_{ab}^T$ against the BOIN boundaries: de-escalation is mandatory if $\hat{p}_{ab}^T > \lambda_d$, escalation is mandatory if $\hat{p}_{ab}^T \le \lambda_e$, and the design otherwise remains within the current toxicity-equivalence zone, refining its position among {stay, de-escalate} once $n_{ab} \ge N^* = 6$ patients have accrued at the cell. Within a mandatory direction, and when refining inside the stay zone, the two (or three) eligible neighbouring cells are ranked by the in-trial posterior exceedance probability $\Pr(u_{ab}^* > u_b^* \mid \mathcal{D}_{ab})$ of Section 3, which reduces to a fixed prior probability for an as-yet-untested cell and therefore does not foreclose exploration. At the maximum sample size, the OBDC is the admissible tested cell (Equation 1, evaluated under the Beta–Binomial posterior) with the largest posterior mean utility.

## 4.3 BHUC: a hierarchical Bayesian model-assisted design for cross-indication OBDC finding

When a combination regimen is developed concurrently in two or more indications, formally sharing information across indications can substantially reduce the sample size needed to identify each indication's OBDC, provided that non-exchangeable information is automatically discounted rather than blindly pooled (Schmidli et al. 2014). We propose the Bayesian Hierarchical Utility-based Cross-indication (BHUC) design, which retains the model-assisted mandatory-direction structure of Section 4.2 independently within each indication (so that safety decisions are always driven by that indication's own toxicity data) but replaces the utility statistic used to rank candidate cells within a direction, or to refine within the toxicity-equivalence zone, with a posterior mean utility computed under a *robust two-component mixture prior* at each cell. Writing $x_{ab}$ and $x^o_{ab}$ for the continuised utility counts (Section 3) in the indication of interest and in the other indication, respectively, the mixture places prior weight one half on a vague, own-data-only $\mathrm{Beta}(a_0, b_0)$ component and one half on a *discounted-pooled* component $\mathrm{Beta}\big(a_0 + \gamma x^o_{ab},\, b_0 + \gamma(n^o_{ab} - x^o_{ab})\big)$, where the fixed discount factor $\gamma \in (0,1)$ down-weights the information borrowed from the other indication in the spirit of a power prior (Schmidli et al. 2014). The posterior weight on each component is updated via its Beta–Binomial marginal likelihood, so that a cell at which the two indications' observed utilities disagree markedly automatically shifts posterior weight onto the vague, own-data-only component—exactly the robustification mechanism that protects against harmful borrowing when indications are not exchangeable, without requiring the analyst to specify in advance which indications will turn out to be concordant. Proposition 4.1 below formalizes this claim: in the large-sample limit at a single cell, the posterior weight on the pooled component is a unimodal function of the own indication's true rate that is maximized when it agrees with the other indication's observed rate and provably vanishes as the two indications' rates become maximally discordant.

**Proposition 4.1** (Automatic discounting under indication discordance)**.** *Fix $a_0, b_0 > 0$, $\gamma \in (0,1)$, and other-indication cell data $(n^o_{ab}, x^o_{ab})$ with $0 < x^o_{ab} < n^o_{ab}$, and let $\alpha_1 = a_0 + \gamma x^o_{ab}$, $\beta_1 = b_0 + \gamma(n^o_{ab} - x^o_{ab})$ denote the resulting discounted-pooled hyperparameters. As own-indication data accrue at cell $(a,b)$ with $x_{ab}/n_{ab} \to \theta^*_{ab} \in (0,1)$ almost surely as $n_{ab} \to \infty$, the Bayes factor comparing the discounted-pooled component to the vague component,*

$$\mathrm{BF}_{10}(\mathcal{D}_{ab}) = \frac{B(x_{ab} + \alpha_1,\; n_{ab} - x_{ab} + \beta_1)\, B(a_0, b_0)}{B(x_{ab} + a_0,\; n_{ab} - x_{ab} + b_0)\, B(\alpha_1, \beta_1)}, \tag{3}$$

*satisfies $\log \mathrm{BF}_{10}(\mathcal{D}_{ab}) \to \Delta(\theta^*_{ab})$ almost surely, where*

$$\Delta(\theta) = \gamma x^o_{ab} \log\theta + \gamma(n^o_{ab} - x^o_{ab}) \log(1-\theta) + \log\frac{B(a_0, b_0)}{B(\alpha_1, \beta_1)}. \tag{4}$$

*The function $\Delta$ is strictly concave on $(0,1)$, uniquely maximized at $\theta = x^o_{ab}/n^o_{ab}$, and satisfies $\Delta(\theta) \to -\infty$ as $\theta \to 0^+$ or $\theta \to 1^-$. Consequently the limiting posterior weight on the pooled component, $\pi^\infty_1(\theta^*_{ab}) = \{1 + \exp(-\Delta(\theta^*_{ab}))\}^{-1}$, is a unimodal function of the own indication's true cell rate that is maximized when it coincides with the other indication's observed rate $x^o_{ab}/n^o_{ab}$ and vanishes as $\theta^*_{ab} \to 0$ or $1$.*

*Proof.* Write $\mathrm{BF}_{10} = \exp\{f(\alpha_1, \beta_1) - f(a_0, b_0)\}$ with

$$f(\alpha, \beta) = \log\Gamma(x_{ab} + \alpha) - \log\Gamma(\alpha)$$

$$
\begin{aligned}
&+ \log\Gamma(n_{ab} - x_{ab} + \beta) - \log\Gamma(\beta) \\
&- \log\Gamma(n_{ab} + \alpha + \beta) + \log\Gamma(\alpha + \beta).
\end{aligned}
$$

Group the six terms of $f(\alpha_1, \beta_1) - f(a_0, b_0)$ into three differences sharing a common, diverging first argument,

$$
\begin{aligned}
f(\alpha_1, \beta_1) - f(a_0, b_0) = &\Big[\log\Gamma(x_{ab} + \alpha_1) \\
&\qquad - \log\Gamma(x_{ab} + a_0)\Big] \\
&+ \Big[\log\Gamma(n_{ab} - x_{ab} + \beta_1) \\
&\qquad - \log\Gamma(n_{ab} - x_{ab} + b_0)\Big] \\
&- \Big[\log\Gamma(n_{ab} + \alpha_1 + \beta_1) \\
&\qquad - \log\Gamma(n_{ab} + a_0 + b_0)\Big] \\
&+ \log\frac{B(a_0, b_0)}{B(\alpha_1, \beta_1)},
\end{aligned}
$$

where the last term collects the three $n_{ab}$-free differences $-[\log\Gamma(\alpha_1) - \log\Gamma(a_0)] - [\log\Gamma(\beta_1) - \log\Gamma(b_0)] + [\log\Gamma(\alpha_1 + \beta_1) - \log\Gamma(a_0 + b_0)] = \log B(a_0, b_0) - \log B(\alpha_1, \beta_1)$, using $B(\alpha, \beta) = \Gamma(\alpha)\Gamma(\beta)/\Gamma(\alpha + \beta)$. By the standard Gamma-function asymptotic $\log\Gamma(z + c) - \log\Gamma(z + c') = (c - c')\log z + O(1/z)$ as $z \to \infty$ (obtained from Stirling's series for $\log\Gamma$), applied with $z = x_{ab}$, $z = n_{ab} - x_{ab}$, and $z = n_{ab}$ respectively,

$$
\begin{aligned}
f(\alpha_1, \beta_1) - f(a_0, b_0) = &(\alpha_1 - a_0)\log x_{ab} \\
&+ (\beta_1 - b_0)\log(n_{ab} - x_{ab}) \\
&- [(\alpha_1 - a_0) + (\beta_1 - b_0)]\log n_{ab} \\
&+ \log\frac{B(a_0, b_0)}{B(\alpha_1, \beta_1)} \\
&+ O(1/x_{ab}) + O(1/(n_{ab} - x_{ab})) \\
&+ O(1/n_{ab}),
\end{aligned}
$$

using $\alpha_1 + \beta_1 - a_0 - b_0 = (\alpha_1 - a_0) + (\beta_1 - b_0)$ for the third bracket. Writing $x_{ab} = n_{ab}\hat{\theta}$ with $\hat{\theta} \to \theta^*_{ab} \in (0, 1)$ almost surely as $n_{ab} \to \infty$ (so that $x_{ab}$, $n_{ab} - x_{ab}$, and $n_{ab}$ all diverge, and the three remainder terms are jointly $O(1/n_{ab})$ since $\hat{\theta}$ is bounded away from 0 and 1 in the limit), substitute $\log x_{ab} = \log n_{ab} + \log\hat{\theta}$ and $\log(n_{ab} - x_{ab}) = \log n_{ab} + \log(1 - \hat{\theta})$ into the bracketed expression:

$$
\begin{aligned}
&(\alpha_1 - a_0)[\log n_{ab} + \log\hat{\theta}] \\
&\quad + (\beta_1 - b_0)[\log n_{ab} + \log(1 - \hat{\theta})] \\
&\quad - [(\alpha_1 - a_0) + (\beta_1 - b_0)]\log n_{ab} \\
&= (\alpha_1 - a_0)\log\hat{\theta} + (\beta_1 - b_0)\log(1 - \hat{\theta}),
\end{aligned}
$$

where the two $\log n_{ab}$ contributions cancel *exactly*, term-for-term, leaving no residual dependence on $n_{ab}$ in the leading order. Hence

$$
f(\alpha_1, \beta_1) - f(a_0, b_0) = (\alpha_1 - a_0)\log\hat{\theta}
$$

$$
\begin{aligned}
&+ (\beta_1 - b_0)\log(1-\hat{\theta}) \\
&+ \log\frac{B(a_0,b_0)}{B(\alpha_1,\beta_1)} + O(1/n_{ab}).
\end{aligned}
$$

Substituting $\alpha_1 - a_0 = \gamma x^o_{ab}$, $\beta_1 - b_0 = \gamma(n^o_{ab} - x^o_{ab})$ and letting $n_{ab} \to \infty$ (so $\hat{\theta} \to \theta^*_{ab}$ almost surely by the strong law of large numbers, and the $O(1/n_{ab})$ remainder vanishes) gives $\log \mathrm{BF}_{10}(\mathcal{D}_{ab}) \to \Delta(\theta^*_{ab})$ as in Equation (4). Differentiating, $\Delta'(\theta) = \gamma x^o_{ab}/\theta - \gamma(n^o_{ab} - x^o_{ab})/(1-\theta)$ and $\Delta''(\theta) = -\gamma x^o_{ab}/\theta^2 - \gamma(n^o_{ab} - x^o_{ab})/(1-\theta)^2 < 0$ whenever $0 < x^o_{ab} < n^o_{ab}$, so $\Delta$ is strictly concave; setting $\Delta'(\theta) = 0$ gives $\theta = x^o_{ab}/n^o_{ab}$. Since $\gamma x^o_{ab} > 0$ and $\gamma(n^o_{ab} - x^o_{ab}) > 0$, each term of $\Delta$ diverges to $-\infty$ as its argument approaches the corresponding boundary, so $\Delta(\theta) \to -\infty$ as $\theta \to 0^+$ or $\theta \to 1^-$. The stated limit for $\pi_1^\infty$ follows from $\pi_1(\mathcal{D}_{ab}) = \mathrm{BF}_{10}/(1+\mathrm{BF}_{10})$ (equal prior component weights) and continuity of $x \mapsto x/(1+x)$. □

(Analogous, weaker statements hold at the boundary cases $x^o_{ab} \in \{0, n^o_{ab}\}$, omitted for brevity, in which the corresponding term of $\Delta$ vanishes identically rather than diverging.) Proposition 4.1 confirms, in the large-sample limit at a single cell, the robustification behaviour illustrated numerically in Section 7: the further the other indication's observed rate departs from the own indication's true rate, the more strongly BHUC's mixture concentrates posterior weight on the own-data-only component, automatically and without the analyst ever declaring in advance which indications are concordant.

Proposition 4.1 is an asymptotic ($n_{ab} \to \infty$) statement, and its rate of approach to the limiting weight $\pi_1^\infty$ depends on $\gamma$ and the amount of other-indication information $n^o_{ab}$; it therefore does not, by itself, bound how much BHUC could rely on borrowed information at the finite, modest per-indication sample sizes used in Sections 6–7. Proposition 4.2 complements it with an exact, sample-size-free guarantee that holds uniformly over every possible own-indication outcome, including at a cell's very first cohort.

**Proposition 4.2** (Finite-sample ceiling on borrowed influence)**.** *For fixed $a_0, b_0, \gamma, n^o_{ab}, x^o_{ab}$ with $0 < x^o_{ab} < n^o_{ab}$, the Bayes factor of Equation (3) satisfies*

$$
\mathrm{BF}_{10}(\mathcal{D}_{ab}) \le M := \left(\frac{x^o_{ab}}{n^o_{ab}}\right)^{\gamma x^o_{ab}} \left(1 - \frac{x^o_{ab}}{n^o_{ab}}\right)^{\gamma(n^o_{ab} - x^o_{ab})} \times \frac{B(a_0,b_0)}{B(\alpha_1,\beta_1)} \tag{5}
$$

*for every $n_{ab} \ge 0$ and every $x_{ab} \in \{0,\dots,n_{ab}\}$—not only in the limit $n_{ab} \to \infty$—so that the posterior weight on the pooled component satisfies $\pi_1(\mathcal{D}_{ab}) \le M/(1+M)$ uniformly over all possible own-indication data, including at an as-yet-untested cell.*

*Proof.* Write $\mathrm{BF}_{10}(\mathcal{D}_{ab}) = \int h(\theta) g_1(\theta)\, d\theta \big/ \int h(\theta) g_0(\theta)\, d\theta$, where $h(\theta) = \theta^{x_{ab}}(1-\theta)^{n_{ab}-x_{ab}}$ is the (nonnegative) binomial kernel and $g_1 = \mathrm{Beta}(\cdot\,; \alpha_1, \beta_1)$, $g_0 = \mathrm{Beta}(\cdot\,; a_0, b_0)$ are the two component densities. For any nonnegative $h$ and any two densities $g_1, g_0$ with $g_0 > 0$ on $(0,1)$,

$$
\begin{aligned}
\int h g_1 \, d\theta &= \int h g_0 \cdot \frac{g_1}{g_0}\, d\theta \\
&\le \left(\sup_{\theta\in(0,1)} \frac{g_1(\theta)}{g_0(\theta)}\right) \int h g_0 \, d\theta,
\end{aligned}
$$

so $\mathrm{BF}_{10}(\mathcal{D}_{ab}) \le \sup_\theta\{g_1(\theta)/g_0(\theta)\}$ for every $n_{ab}$ and $x_{ab}$, with no asymptotics required. Now

$$\begin{aligned}\frac{g_1(\theta)}{g_0(\theta)} &= \frac{B(a_0,b_0)}{B(\alpha_1,\beta_1)}\,\theta^{\alpha_1-a_0}(1-\theta)^{\beta_1-b_0}\\ &= \frac{B(a_0,b_0)}{B(\alpha_1,\beta_1)}\,\theta^{\gamma x^o_{ab}}(1-\theta)^{\gamma(n^o_{ab}-x^o_{ab})},\end{aligned}$$

which, by the same calculation as in the proof of Proposition 4.1, is uniquely maximized at $\theta = x^o_{ab}/n^o_{ab}$; substituting gives $M$ in Equation (5). The weight bound follows from $\pi_1 = \mathrm{BF}_{10}/(1+\mathrm{BF}_{10})$ and monotonicity of $x \mapsto x/(1+x)$. □

To make the gap between Propositions 4.1 and 4.2 concrete, consider an illustrative discordant cell with $a_0 = b_0 = 1$, $\gamma = 0.5$, and other-indication data $n^o_{ab} = 30$, $x^o_{ab} = 12$ (observed rate 0.40)—representative of the $N = 30$-per-indication design used throughout this article—at which the own indication's true rate is $\theta^*_{ab} = 0.70$. The finite-sample ceiling of Proposition 4.2 gives $\pi_1(\mathcal{D}_{ab}) \le 0.77$, holding regardless of $n_{ab}$; the large-sample limit of Proposition 4.1 is $\pi_1^\infty = 0.16$; and at the trial's actual per-indication sample size, $n_{ab} = 30$, direct numerical evaluation of Equation (3) gives $\pi_1(\mathcal{D}_{ab}) \approx 0.31$—below the ceiling, substantially reduced from the untested-cell value of 0.50 (Remark below), but still well short of the asymptotic value, since convergence in $n_{ab}$ is genuinely slow for this discount and this amount of other-indication information. We report this honestly rather than presenting Proposition 4.1 as a finite-sample guarantee: the robustification mechanism is real and moves in the right direction well before $n_{ab} \to \infty$, as the simulation and sensitivity results of Sections 6–7 confirm empirically at $N = 30$, but it is not yet close to its asymptotic strength at the sample sizes typical of a single early-phase indication, and a trial team relying on BHUC should treat the discounting as a directionally reliable, partial safeguard rather than a near-complete one at these sample sizes.

**Remark 4.3** (Behaviour at an untested cell)**.** At a cell with no own-indication data ($n_{ab} = 0$), the marginal likelihood of "zero observations" under a Beta–Binomial model is exactly 1 under either mixture component, for any hyperparameters, so $\mathrm{BF}_{10}(\mathcal{D}_{ab}) = 1$ and $\pi_1(\mathcal{D}_{ab}) = 0.5$ identically: with no own evidence to favour either component, BHUC's utility ranking at that cell is an equally weighted blend of the vague own-indication prior mean and the discounted-pooled (other-indication) prior mean, rather than either a full commitment to borrowed information or a full disregard of it. The discordance-driven down-weighting of Proposition 4.1 therefore only begins to operate from a cell's *second* cohort onward, once $n_{ab} > 0$; a cell's first cohort is necessarily allocated under this 50/50 blend, since no design—rule-based, model-assisted, or model-based—can adapt to evidence it does not yet have. In a real combination trial developed across indications, this means the very first patients assigned to a newly opened cell in a new indication are exposed to a ranking statistic that is already half-anchored to whatever the other indication's data show at that cell, before any of the new indication's own safety or efficacy data exist to check it; if the indications later prove discordant, this initial reliance is corrected automatically as own data accrue (Proposition 4.1), but only gradually, and Proposition 4.2 shows the correction can take materially more than the handful of patients accrued in a typical early-phase cohort to approach its asymptotic strength. We view this as an inherent, honestly disclosed limitation of hierarchical borrowing at the moment before indication-specific evidence exists, common to any Bayesian borrowing design and not specific to BHUC's particular mixture construction, rather than a defect that a different weighting scheme would remove.

We adopt this robust mixture, rather than a fully exchangeable hierarchical random-effects model with a shared normal or logit-normal random effect across indications, as the "best" hierarchical construction for this problem for three practical reasons: (i) with only two or a handful of indications and $J \times K = 16$ cells each, a random-effects variance component is weakly identified and its posterior is highly sensitive to its own hyperprior, whereas the two-component mixture requires no variance component to be estimated; (ii) the mixture weight is cell-specific and recomputed at every interim decision in closed form, avoiding Markov chain Monte Carlo and so preserving the model-assisted design's computational transparency; and (iii) the mixture degrades gracefully to no borrowing (weight concentrating on the vague component) at any discordant cell while a random-effects model borrows everywhere by the same, globally estimated amount. Terminal OBDC selection in each indication proceeds exactly as in Section 4.2, substituting the robust mixture posterior mean for the single-source posterior mean.

### *4.3.1 Relationship to existing cross-trial borrowing frameworks*

BHUC's discounted-pooled-versus-vague mixture is not proposed in a vacuum, and it is useful to place it explicitly against the three established families of Bayesian borrowing methods with which a referee familiar with this literature would compare it: fixed power priors (Chen and Ibrahim 2000), commensurate priors (Hobbs et al. 2011), and robust meta-analytic-predictive (robust MAP) mixture priors (Schmidli et al. 2014), the last of which is the direct methodological ancestor of BHUC's construction (see also the review of historical- and concurrent-borrowing frameworks by Viele et al. (2014)). A fixed power prior (Chen and Ibrahim 2000) raises the other source's likelihood contribution to a pre-specified power $\gamma \in (0, 1)$ and pools it with the current data unconditionally: it is exactly BHUC's discounted-pooled component, Equation (3)'s component 2, used on its own, with no vague-component alternative and therefore no mechanism for the discount actually applied to adapt to how well the two sources agree. A commensurate prior (Hobbs et al. 2011) instead lets the current-indication mean shrink toward the other indication's mean through a continuous commensurability parameter that is itself estimated from the data (typically via a normal random-effect linking the two indication-specific means, with its own precision hyperparameter learned at each analysis), so that, like BHUC, the effective amount of borrowing is data-driven rather than fixed; unlike BHUC, that estimation is continuous and requires either numerical integration or Markov chain Monte Carlo to update at every interim decision, cell by cell across the $J \times K$ grid, which conflicts with the closed-form, MCMC-free updating that keeps BHUC (and Comb-BOIN12, from which it inherits its quasi-binomial machinery) tractable for real-time dose-finding decisions and for a 5000-replication simulation study. A robust MAP prior (Schmidli et al. 2014) is architecturally closest to BHUC: both use a two-component mixture of an informative, externally-pooled component and a vague, own-data-only component, with the posterior mixture weight itself acting as the data-driven discordance check formalized in Proposition 4.1. The difference is one of application, not construction: Schmidli et al. (2014) and Viele et al. (2014) develop and review this mixture for borrowing from a fixed, already-observed historical-control dataset into a single current trial, whereas BHUC applies the identical mechanism reciprocally and concurrently between two indications that are both still accruing within the same combination trial, cell by cell across the dose grid, with a pre-fixed discount $\gamma$ multiplying the Beta–Binomial pooled component rather than a continuously re-estimated weight – a deliberate simplification, justified by reasons (i)–(iii) above, that keeps the mixture weight of Equation (3) available in closed form at every interim look.

To make the comparison to a fixed (non-robust) power prior concrete rather than only conceptual, we implemented a design identical to BHUC in every respect — the same discount $\gamma = 0.5$, the same sequential dose-movement rule, the same admissibility criteria, the same terminal selection

rule — except that the utility statistic always uses the single discounted-pooled Beta posterior of component 2, with no vague component and no discordance-driven down-weighting, and reran the concordant- and discordant-pair simulations of Section 6 against it (Table 5). Under the concordant pair, the two methods perform comparably, as expected: with true rates that agree, there is nothing for robustification to correct, and the fixed power prior borrows the same useful information BHUC's mixture also converges toward. Under the discordant pair, at this discount and this per-indication sample size ($N = 30$), the fixed power prior's correct-selection percentages are statistically indistinguishable from BHUC's in Indication 2 and modestly higher in Indication 1 (Table 5); repeating the comparison at a substantially larger discount, $\gamma = 0.9$, leaves this picture essentially unchanged (Supplementary Material, Section S7). We report this without qualification because it is the honest empirical finding, not the one that would make the strongest case for robustification: at the modest discordance and sample size realized in our two-indication scenarios, the two methods' average-case operating characteristics are close. What the fixed power prior cannot offer, and what BHUC's mixture provides by construction, is a worst-case guarantee that is entirely independent of how discordant the two indications happen to be in any particular simulation: Proposition 4.2's finite-sample ceiling holds for every possible own-indication outcome, including a cell's very first cohort, and bounds how far the borrowed component can dominate the posterior mixture regardless of $n_{ab}$. A fixed power prior has no analogue of this bound — its borrowed influence is exactly $\gamma$, unconditionally, for any degree of discordance and any amount of other-indication data $n^o_{ab}$, so a sufficiently adversarial disagreement (larger $\gamma$, larger $n^o_{ab}$, or a starker rate mismatch than either of our two simulated pairs realizes) biases a fixed power prior by an amount that grows without limit, while Proposition 4.2's illustrative calculation, given immediately after its proof in Section 4.3, already shows BHUC's corresponding weight capped at 0.77 under comparable conditions. We therefore recommend the robust mixture not because it necessarily wins on correct selection in every trial one might simulate, but because it is the only one of the four frameworks compared here that provides a distribution-free, sample-size-free bound on the damage indication-specific borrowing can do, a property a fixed power prior and (without additional, MCMC-dependent machinery) a commensurate prior do not share.

### 4.4 EffTox-approx: model-based benchmark

We benchmark against a model-based design in the spirit of EffTox (Thall and Cook 2004; Brock et al. 2017), using independent logistic regressions for toxicity and efficacy on standardized dose scores $\tilde{a}_j, \tilde{b}_k \in \{-0.58, -0.19, 0.19, 0.58\}$ (mean zero, chosen to avoid the collinear interaction term $\log(d_{1j}d_{2k}) = \log d_{1j} + \log d_{2k}$ that arises when raw log-doses are used with an additive-plus-product linear predictor),

$$\begin{aligned} \operatorname{logit} p^T_{jk} &= \beta^T_0 + \beta^T_1 \tilde{a}_j + \beta^T_2 \tilde{b}_k + \beta^T_3 \tilde{a}_j \tilde{b}_k, \\ \operatorname{logit} p^E_{jk} &= \beta^E_0 + \beta^E_1 \tilde{a}_j + \beta^E_2 \tilde{b}_k + \beta^E_3 \tilde{a}_j \tilde{b}_k, \end{aligned} \tag{6}$$

fitted at every interim decision by penalized (ridge) Newton–Raphson maximum a posteriori estimation with a Laplace approximation to the posterior covariance, avoiding Markov chain Monte Carlo so that 5000 replications of a multi-scenario simulation study remain computationally tractable; posterior draws from the resulting multivariate normal approximation are evaluated simultaneously across the whole $J \times K$ grid at each decision (Supplementary Material, Section S2) to make the approximation fast enough for large-scale simulation without altering its statistical logic. Escalation follows the neighbouring-cell rule of Thall and Cook (2004): among admissible neighbours of the current cell, the design moves to the one with highest posterior mean utility, or de-escalates toward the origin if no cell is currently admissible. We report this design's genuine operating characteristics, including its known sensitivity to sparse data in small grids (Section 8), rather than presenting only favourable scenarios, so that the roadmap in Section 5 reflects real trade-offs rather than an idealized comparison.

## 5 A Practical Roadmap for Design Selection

Figure 2 distils Sections 2 and 4, together with the operating characteristics in Sections 6–8, into a decision aid for a statistician planning a two-agent combination trial. Three questions suffice to select a design in the large majority of cases. First, is the regimen being developed in more than one indication with a shared mechanism of action, such that borrowing information across indications is scientifically defensible? If so, BHUC (Section 4.3) should be preferred over running independent per-indication designs, since Section 6 shows it improves correct selection even when indications are discordant. Second, does the trial team require that the complete decision rule be auditable and tabulable in the protocol without reference to a fitted statistical model, for example because of regulatory or site-level operational constraints? If so, a rule-based design (Ji3+3-Comb) is appropriate, at the cost of the higher "none-selected" rates documented in Section 6 under flat or plateauing efficacy surfaces. Third, if neither of the above applies, is the maximum planned sample size comfortably above roughly thirty patients per dose-finding cohort, with dedicated statistical support available to monitor a fitted model in real time? If so, a model-based design such as EffTox-approx may be used, understanding that it can be less stable than a model-assisted design at smaller sample sizes (Section 8); if not, a model-assisted design (Comb-BOIN12, or BHUC when multiple indications are involved) offers the most robust safety–efficacy trade-off across the scenarios we examined and is our default recommendation.

## 6 Simulation Study

### 6.1 Design

We evaluated Ji3+3-Comb, Comb-BOIN12, and EffTox-approx across six single-indication scenarios (Table 2) spanning a monotone dose–response surface informed by the dose ranges of the MATCHPOINT trial (Brock et al. 2017), an efficacy plateau informed by the dose ranges of CheckMate 012 (Hellmann et al. 2017), an interior efficacy "ridge/valley" surface reflecting antagonism at intermediate dose ratios, a high-background-toxicity surface in which only the lowest corner of the grid is safe, a scenario in which every cell exceeds $\phi_T$ (so that the correct action is to declare no acceptable dose), and a scenario in which the OBDC coincides with the maximum tolerated dose combination. All scenarios use a $4 \times 4$ grid, cohorts of three patients, a maximum sample size of $N = 36$, $\phi_T = 0.35$, $\phi_E = 0.20$, $w_T = 0.5$, and $C_T = C_E = 0.90$. Each design–scenario combination was simulated for 5000 replications. We separately evaluated BHUC against Comb-BOIN12 run independently in each indication ("no borrowing"), on a concordant pair of indications sharing the same underlying surface and a discordant pair whose true OBDCs lie in different regions of the grid, with $N = 30$ patients per indication and $\gamma = 0.5$; full R code is given in the Supplementary Material.

### 6.2 Results

Table 3 and Figure 3 summarize the single-indication results. No single design dominates: Comb-BOIN12 gives the most consistent correct-selection percentage across scenarios with moderate overdose risk; Ji3+3-Comb is competitive under S3 and safest under S4–S5 but frequently declares no dose ("none-selected") under S1, S2, and S4, reflecting the conservatism of comparing raw observed proportions with no model-based smoothing; EffTox-approx performs best when the OBDC coincides with the maximum dose (S6) but shows markedly elevated overdose selection under S1, S2, and especially S5, where every cell is truly unsafe—a known vulnerability of parametric extrapolation from sparse interior data that we revisit in Section 8. Under S1 and S2, correct-selection percentages are modest for all three designs; Section 6.3 below shows that this is driven substantially, though not entirely, by the closely spaced utilities of several admissible cells under these surfaces (Supplementary Material, Section S1), so we also report the percentage of patients allocated to the true OBDC (column PatOBDC) as a complementary operating characteristic. Under S5, where no dose is truly acceptable, Ji3+3-Comb correctly declares no dose in 94.5% of trials, versus 39.1% for EffTox-approx and 18.4% for Comb-BOIN12, illustrating a genuine and scenario-dependent safety–exploration trade-off between the three paradigms.

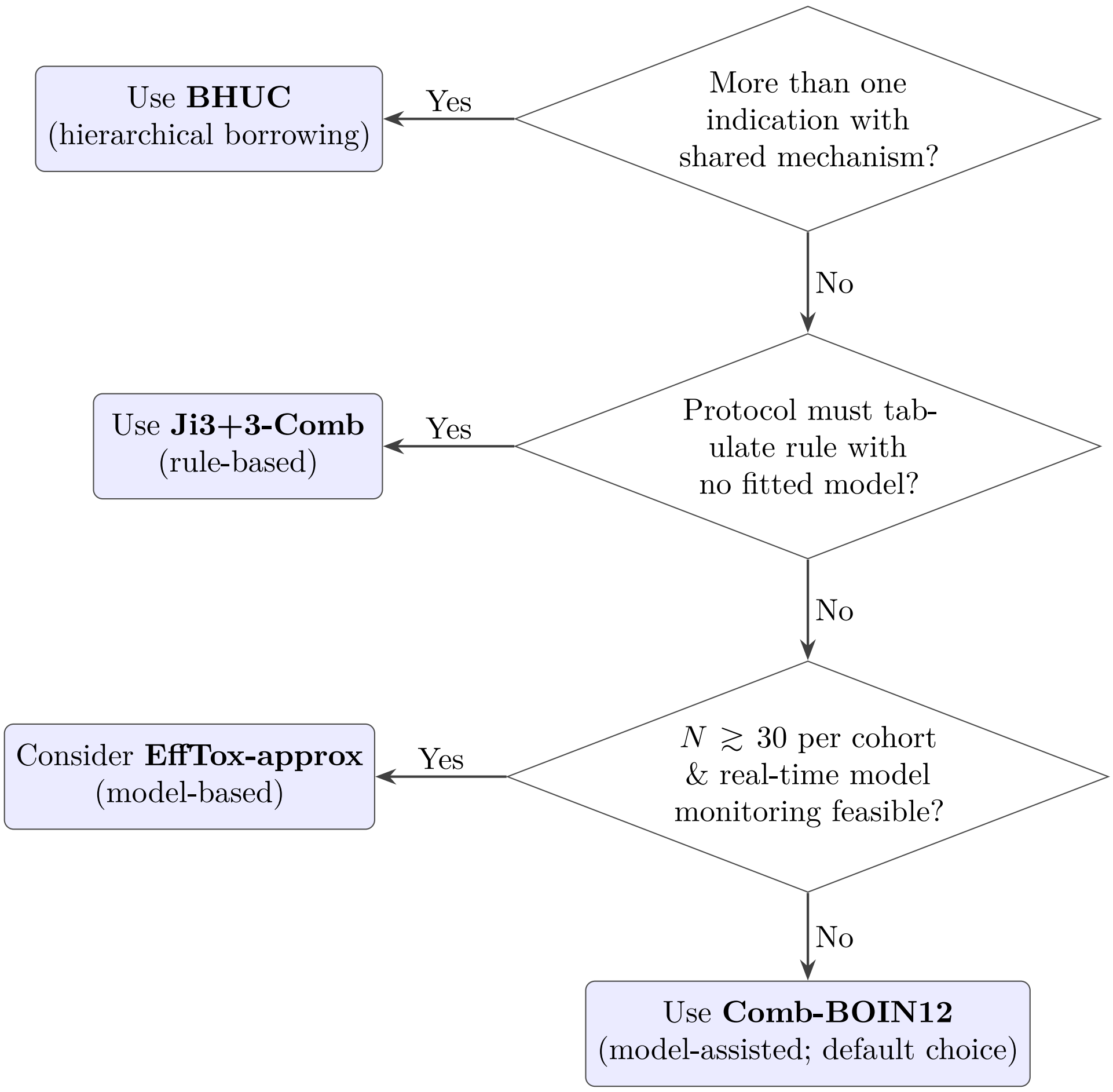

**Figure 2** Practical roadmap for selecting a Bayesian OBDC design, based on the operating characteristics reported in Sections 6–8.

### 6.3 The achievable ceiling on correct selection under closely-spaced utilities

To assess how much of the modest correct-selection percentages under S1–S4 is attributable to the sampling noise inherent in distinguishing closely spaced true utilities at $N = 36$, rather than to a shortcoming of the three designs themselves, we computed an oracle upper bound: a hypothetical procedure that is handed the *true* admissible set in advance (so it can neither wrongly exclude the true OBDC nor wrongly include an inadmissible cell) and splits the entire $N = 36$ budget as evenly as possible among just those cells, applying the same $\text{Beta}(1, 1)$ posterior-mean terminal rule used by Comb-BOIN12. Equal allocation to a known candidate set is the classical optimal design for this pure ranking-and-selection problem, so no nonparametric, per-cell procedure — rule-based or model-assisted, including Ji3+3-Comb and Comb-BOIN12 — that must additionally discover the admissible set and traverse the dose grid via a constrained escalation path can exceed it with the same total $N$. The bound does not apply to EffTox-approx, whose parametric model borrows strength across the whole grid rather than replicating within single cells, and it is not meaningful under S6, where 13 of 16 cells are jointly admissible and a sequential design's natural concentration near the safe, high-dose region lets it substantially outperform a naive equal split; we therefore report the ceiling only for S1–S4, where it is both valid and informative (Supplementary Material, Section S7).

**Table 2** Single-indication simulation scenarios and the true OBDC cell (dose level of agent 1, dose level of agent 2) under each; "—" denotes that no cell is admissible.

| Scen. | Description | Motivation | True OBDC |
|---|---|---|---|
| S1 | Monotone efficacy & toxicity | MATCHPOINT-scale targeted-agent doublet (Brock et al. 2017) | $(3, 3)$ |
| S2 | Efficacy plateau | CheckMate 012-scale immunotherapy doublet (Hellmann et al. 2017) | $(3, 3)$ |
| S3 | Interior ridge/valley | Antagonism at intermediate dose ratios | $(4, 1)$ |
| S4 | High background toxicity | Only the lowest corner is safe | $(2, 2)$ |
| S5 | All cells overly toxic | No acceptable dose exists | — |
| S6 | OBDC = MTDC | Low toxicity throughout, monotone efficacy | $(4, 4)$ |

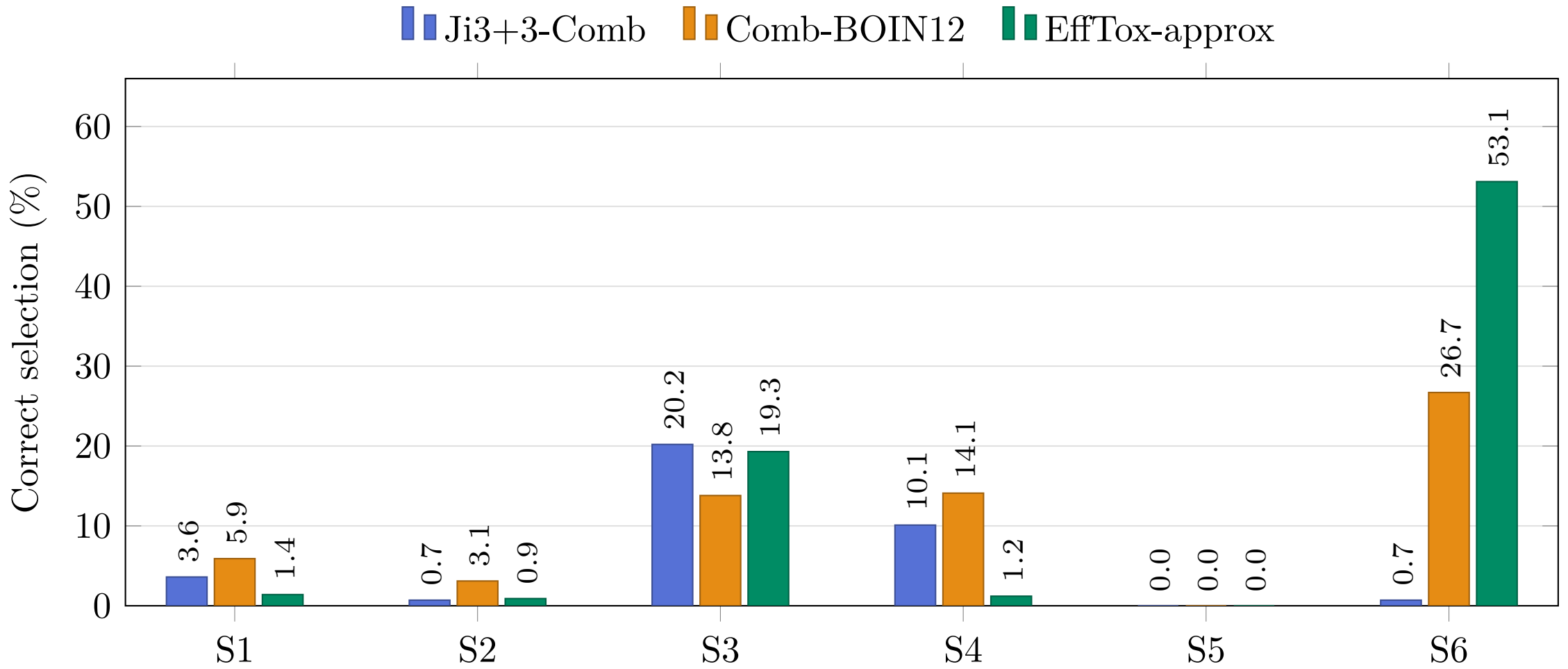

**Figure 3** Percentage of trials correctly selecting the true OBDC, by design and scenario (S5 is undefined, since no dose is truly admissible, and is shown as zero). Bars are grouped by scenario; see Table 3 for the corresponding safety metrics.

Table 4 shows that the ceiling itself is well below 100% under every scenario — confirming that some shortfall from perfect selection is a genuine, unavoidable consequence of these surfaces' closely spaced utilities at $N = 36$, not merely a design artifact — but also that the best-performing design in each scenario achieves only 25–59% of this ceiling, leaving a real, honestly acknowledged gap rather than a claim that the designs are already statistically optimal. We attribute this residual gap to the additional cost that any *sequential*, path-constrained design pays relative to the oracle: unlike the oracle, a real design must spend part of its budget discovering which

**Table 3** Single-indication simulation results (5000 replications per cell; $N = 36$). PCS: % correct selection of the true OBDC (undefined under S5, shown as —). PatOBDC: % of patients allocated to the true OBDC. OverdoseSel: % of trials selecting a cell with true $p^T > \phi_T$. NoneSel: % of trials declaring no acceptable dose.

| Scen. | Design | PCS (%) | PatOBDC (%) | OverdoseSel (%) | NoneSel (%) |
|---|---|---|---|---|---|
| S1 | Ji3+3-Comb | 3.6 | 1.2 | 3.3 | 9.1 |
| | Comb-BOIN12 | 5.9 | 3.6 | 17.2 | 0.0 |
| | EffTox-approx | 1.4 | 3.2 | 39.6 | 0.0 |
| S2 | Ji3+3-Comb | 0.7 | 0.2 | 0.6 | 4.8 |
| | Comb-BOIN12 | 3.1 | 1.9 | 11.8 | 0.0 |
| | EffTox-approx | 0.9 | 2.3 | 32.8 | 0.0 |
| S3 | Ji3+3-Comb | 20.2 | 15.2 | 8.3 | 18.7 |
| | Comb-BOIN12 | 13.8 | 14.1 | 12.9 | 0.0 |
| | EffTox-approx | 19.3 | 8.9 | 26.4 | 0.0 |
| S4 | Ji3+3-Comb | 10.1 | 4.9 | 9.9 | 16.5 |
| | Comb-BOIN12 | 14.1 | 10.5 | 23.3 | 0.1 |
| | EffTox-approx | 1.2 | 4.9 | 54.0 | 0.1 |
| S5 | Ji3+3-Comb | — | — | 5.5 | 94.5 |
| | Comb-BOIN12 | — | — | 81.6 | 18.4 |
| | EffTox-approx | — | — | 60.9 | 39.1 |
| S6 | Ji3+3-Comb | 0.7 | 0.3 | 0.0 | 4.5 |
| | Comb-BOIN12 | 26.7 | 27.0 | 0.0 | 0.0 |
| | EffTox-approx | 53.1 | 23.3 | 0.0 | 0.0 |

Scenario labels S1–S6 refer to Table 2.

cells are admissible, cannot allocate patients directly to the closest competing cells without first reaching them via adjacent-cell escalation, and may allocate replications to cells that turn out not to be among the closest competitors at all. This gap is therefore not a defect specific to any one of the three designs compared here — it is the generic price of on-line, adaptive dose-finding relative to a hypothetical batch procedure with the same total sample size — and it is the same reason we report PatOBDC alongside PCS throughout: a design can accrue most of its patients at or near the true OBDC (high PatOBDC) while still failing, on a given trial, to have its final point estimate land exactly on it (modest PCS), particularly under S1 and S2, where Supplementary Material Section S1 shows the two best admissible cells differ in true utility by as little as 0.005–0.02.

Table 5 and Figure 4 report the cross-indication borrowing study. Both borrowing methods—BHUC and the fixed, non-robust power prior of Section 4.3.1—more than double the correct-selection percentage relative to running Comb-BOIN12 independently in each indication under the concordant pair (Indication 1: 9.7% and 9.3% versus 4.4%; Indication 2: 9.2% and 10.0% versus 3.0%), and continue to improve correct selection under the discordant pair (Indication 1: 10.9% and 12.4% versus 4.2%; Indication 2: 5.9% and 6.4% versus 1.7%), where naive pooling would be expected to harm Indication 2 most. Both borrowing methods therefore borrow usefully relative to no borrowing at all; as discussed in Section 4.3.1, the fixed power prior's correct-selection percentages are, on these two scenarios, comparable to and in the discordant pair modestly higher than BHUC's, so the case for BHUC's added robustification rests on the distribution-free worst-case guarantee of Proposition 4.2, which the fixed power prior cannot offer, rather than on an average-case advantage visible in this particular pair of scenarios. All three designs used the full planned sample size in both indications, so the differences shown are attributable entirely to more efficient use of the same data rather than to differential accrual.

**Table 4** Oracle ceiling on correct selection (equal allocation of $N = 36$ across the true admissible set, 5000 replications) versus the best-performing design's actual correct-selection percentage, under the four scenarios where correct selection is well-defined and per-cell allocation is a meaningful comparison (see text).

| Scen. | Admiss. | Oracle (%) | Best (%) | Ratio (%) |
|---|---|---|---|---|
| S1 | 7 | 20.2 | 5.9 | 29 |
| S2 | 11 | 12.3 | 3.1 | 25 |
| S3 | 4 | 34.3 | 20.2 | 59 |
| S4 | 5 | 26.5 | 14.1 | 53 |

Admiss.: number of admissible cells. Oracle: equal-allocation ceiling. Best: best-performing design's actual PCS (Comb-BOIN12 under S1, S2, and S4; Ji3+3-Comb under S3; see Table 3). Ratio: Best/Oracle.

**Table 5** Cross-indication borrowing results (5000 replications; $N = 30$ per indication). PCS: % correct selection of the true OBDC in each indication. The fixed power prior applies BHUC's same $\gamma = 0.5$ discount to the pooled component unconditionally, with no robust vague component (Section 4.3.1).

| Indication pair | Method | PCS, Ind. 1 (%) | PCS, Ind. 2 (%) |
|---|---|---|---|
| Concordant | Comb-BOIN12 (no borrowing) | 4.4 | 3.0 |
| | Fixed power prior (non-robust) | 9.3 | 10.0 |
| | BHUC (proposed) | 9.7 | 9.2 |
| Discordant | Comb-BOIN12 (no borrowing) | 4.2 | 1.7 |
| | Fixed power prior (non-robust) | 12.4 | 6.4 |
| | BHUC (proposed) | 10.9 | 5.9 |

## 7 Sensitivity Analysis

To assess robustness to prior and model-assumption choices, we repeated key comparisons (2000 replications each) along three axes.

**Borrowing discount factor.** We varied the BHUC discount $\gamma \in \{0.25, 0.5, 0.75\}$ (Table 6). Correct selection is essentially stable across this range for both indications in both pairs, confirming that the robust mixture's automatic down-weighting of discordant information, rather than the precise value of $\gamma$, is what drives BHUC's performance; a trial team need not calibrate $\gamma$ precisely to realize most of the borrowing benefit.

**Utility weight (model assumption).** We varied $w_T \in \{0.3, 0.5, 0.7\}$, recomputing each scenario's true OBDC and true utility surface accordingly (Figure 5). Comb-BOIN12's overdose selection rate falls monotonically as $w_T$ increases (more weight on toxicity avoidance), as expected; Ji3+3-Comb is the most stable design across $w_T$, while EffTox-approx's overdose rate is highest at low $w_T$ and falls sharply as $w_T$ increases, indicating that its safety behaviour is more sensitive to the assumed toxicity penalty than the two model-assisted/rule-based alternatives.

**Quasi-binomial prior informativeness.** We varied the Beta prior $(a_0, b_0) \in \{(0.5, 0.5), (1, 1), (2, 2)\}$ underlying Comb-BOIN12's utility statistic (Table 7). Correct-selection and overdose-selection percentages change by at most one to two percentage points across this range under both S1 and S4, indicating that Comb-BOIN12's operating characteristics are not materially sensitive to the informativeness of the quasi-binomial prior within this range—reassuring given that a trial team rarely has strong grounds to prefer one weakly informative Beta prior over another at the design stage.

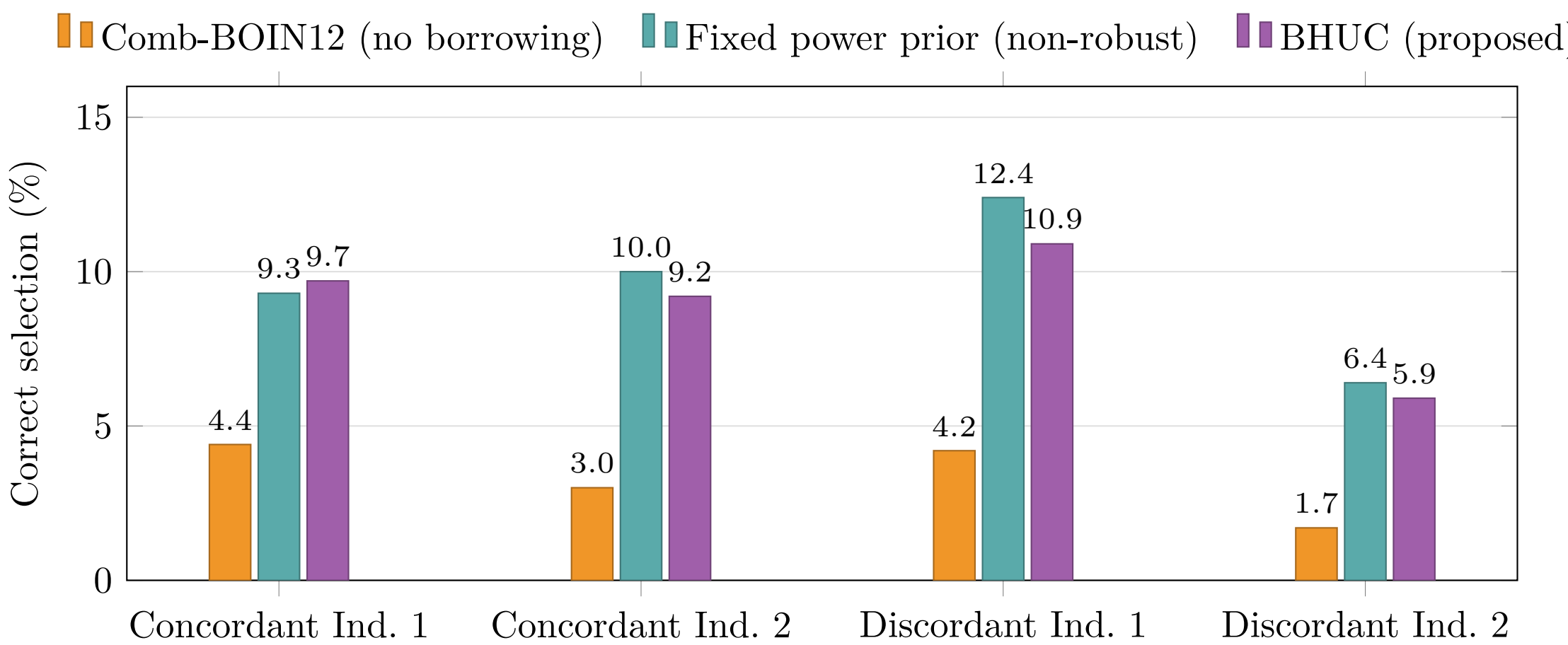

**Figure 4** Correct-selection percentage under no borrowing, fixed (non-robust) power-prior borrowing, and BHUC's robust mixture, for concordant and discordant indication pairs.

**Table 6** Sensitivity of BHUC correct selection (%) to the borrowing discount $\gamma$ (2000 replications).

| | Concordant | | Discordant | |
|---|---|---|---|---|
| $\gamma$ | Ind. 1 | Ind. 2 | Ind. 1 | Ind. 2 |
| 0.25 | 9.6 | 9.1 | 11.1 | 6.2 |
| 0.50 | 11.3 | 10.7 | 12.0 | 5.4 |
| 0.75 | 10.5 | 10.0 | 11.8 | 5.7 |

## 8 Illustrative Case Study

To ground the comparison in real trial data, we constructed a case study from the published dose-escalation results of a phase Ib trial of the WEE1 inhibitor adavosertib combined with the PARP inhibitor olaparib in patients with refractory solid tumours (NCT02511795; Hamilton et al. 2024), which is, to our knowledge, the most completely reported per-cohort two-agent dose-escalation dataset publicly available for a targeted-agent combination. The trial's 15 dose-escalation cohorts tested six adavosertib regimens (125–300 mg, once or twice daily, on several intermittent schedules) against three olaparib doses (100, 200, 300 mg twice daily), enrolling a total of 120 patients with between 3 and 17 patients per cohort and 0–2 dose-limiting toxicities observed per cohort. We collapsed the six adavosertib regimens into three ordinal intensity tiers (Table 8) to obtain a $3 \times 3$ grid directly comparable to our simulation framework, summing patient and toxicity counts for cohorts mapping to the same cell; two cells (lowest adavosertib tier with the highest olaparib dose, and highest adavosertib tier with the lowest olaparib dose) were never tested and their toxicity rates were interpolated under a monotonicity assumption, as flagged in the table. Per-cohort response data were reported in the source publication only for the two candidate recommended-phase-2-dose cohorts (30.8% objective response rate at 175 mg twice-daily adavosertib with 200 mg twice-daily olaparib; 9.1% at 200 mg once-daily adavosertib with 200 mg twice-daily olaparib); efficacy at the remaining cells was not tabulated in the source and is interpolated here, anchored to these two reported rates under a monotone, plateauing dose–efficacy assumption, purely for illustrative purposes—this limitation, itself common to early-phase combination trials that formally assess response only around leading dose candidates, is exactly the kind of evidentiary gap that a prospectively Bayesian-designed OBDC trial is intended to close.

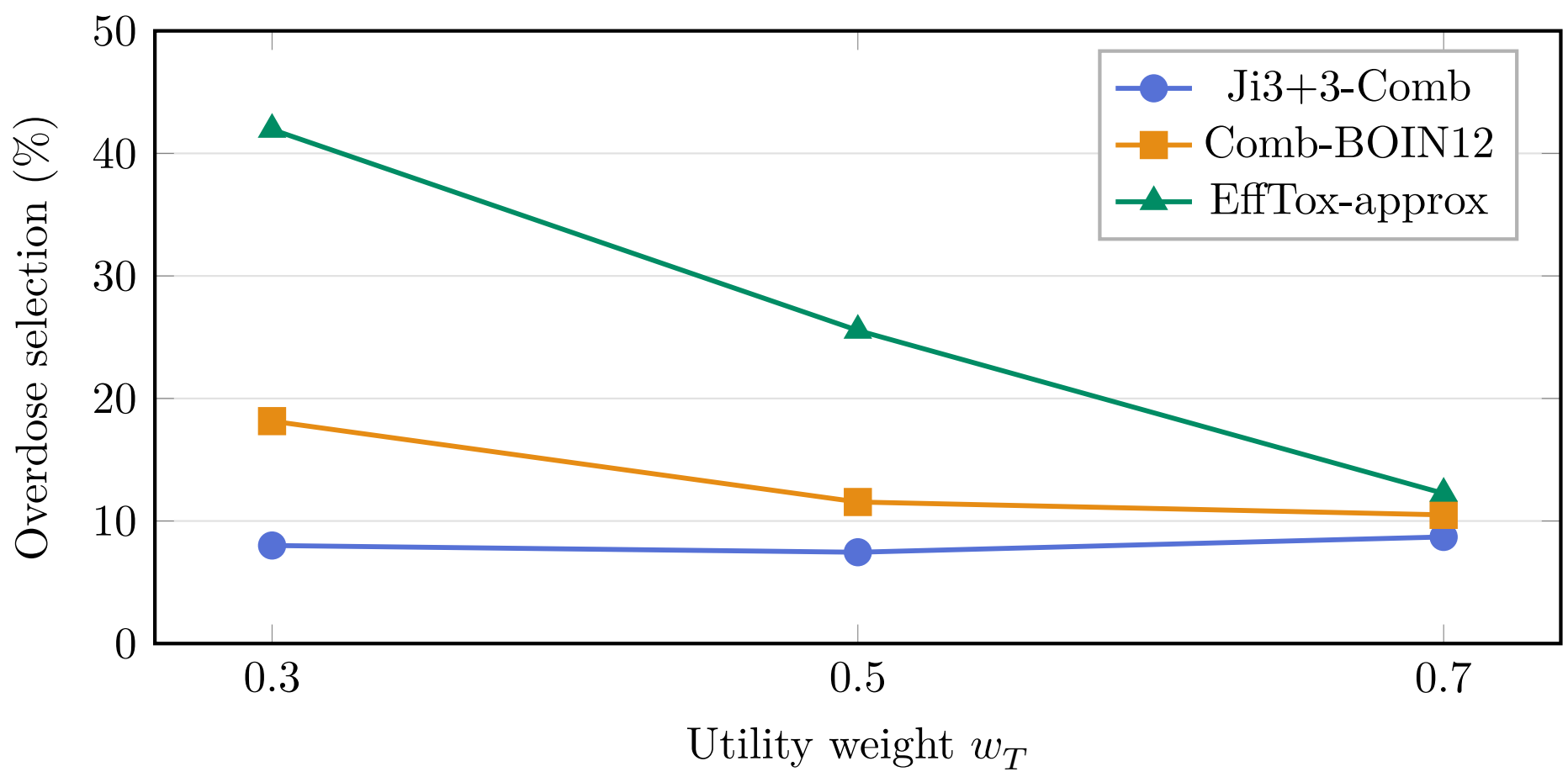

**Figure 5** Overdose-selection percentage under scenario S3 (ridge/valley) as the assumed utility weight $w_T$ varies, holding all other design parameters fixed.

**Table 7** Sensitivity of Comb-BOIN12 operating characteristics (%) to the quasi-binomial prior Beta$(a_0, b_0)$ (2000 replications).

| | S1: monotone | | S4: high tox. | |
|---|---|---|---|---|
| Prior | PCS | OverdSel | PCS | OverdSel |
| Beta$(0.5, 0.5)$ | 6.1 | 18.8 | 13.3 | 23.8 |
| Beta$(1, 1)$ | 5.5 | 18.2 | 12.9 | 23.1 |
| Beta$(2, 2)$ | 5.7 | 18.1 | 13.9 | 22.1 |

Under this assumed truth, the only admissible cells (Equation 1, with $\phi_T = 0.35$, $\phi_E = 0.20$) are (Medium, 200 mg) and (Medium, 300 mg), and the illustrative OBDC is (Medium adavosertib, 200 mg olaparib), with utility 0.265 versus 0.10 at the other admissible cell. This differs from the dose regimen the trial actually carried forward (High adavosertib tier/200 mg olaparib, chosen mainly for the practical convenience of once-daily dosing), illustrating precisely the paradigm shift Project Optimus is intended to drive: a dose chosen principally for tolerability and convenience is not automatically the dose offering the best available efficacy–toxicity trade-off. Replaying each design on this grid for 2000 simulated trials of $N = 36$ patients (Table 9, Figure 6) shows Comb-BOIN12 achieving the highest correct-selection percentage (11.8%) with zero trials declaring no acceptable dose; Ji3+3-Comb frequently drifts to the highest-dose corner (38.1% of trials), reflecting the limitation identified in Section 6 that a purely rule-based design with no smoothing can overshoot when efficacy plateaus rather than increasing monotonically; and EffTox-approx performs poorly (0.1% correct selection), spreading its selections across the grid's corners—direct, real-data-grounded confirmation of the fragility of unregularized joint logistic surfaces at realistic early-phase sample sizes that motivates our general preference for model-assisted designs in Section 5.

## 9 Discussion

This article makes three contributions to the design of early-phase, two-agent combination oncology trials targeting the optimal biological dose combination. First, we give a taxonomy of Bayesian OBDC designs that separates decision mechanism from design objective, resolving the overlapping classifications used informally in the existing literature and providing the scaffolding for the roadmap in Section 5. Second, we introduce Ji3+3-Comb, closing a documented gap by providing a transparent, fully rule-based design for utility-integrated two-agent dose-finding, extending the single-agent Joint i3+3 design (Lin and Ji 2020) to the combination setting. Third, we introduce BHUC, a model-assisted design with a robust mixture-prior hierarchical borrowing mechanism for identifying indication-specific OBDCs when a combination regimen is developed across multiple tumour types, showing across concordant and discordant indication pairs, and across a three-fold range of the borrowing discount, that it improves correct selection without requiring the discount to be precisely calibrated. Both new designs are grounded theoretically rather than posited by analogy: the shared utility function of Section 3 is derived as the Bayes-optimal decision under a linear clinical loss (Proposition 3.1), and BHUC's robust mixture-prior borrowing mechanism is shown to discount information from a discordant indication automatically, with the posterior weight it assigns to pooling provably vanishing in the large-sample limit as the two indications' true rates diverge (Proposition 4.1), complemented by an exact, sample-size-free ceiling on that weight that holds even before any own-indication data accrue (Proposition 4.2); we report explicitly that the asymptotic discounting is not yet complete at the modest per-indication sample sizes used in our simulations, so that the mechanism should be understood as a directionally reliable, partial safeguard at these sample sizes rather than a near-exact one.

**Table 8** Case-study $3 \times 3$ grid derived from Hamilton et al. (2024). Cell entries are (patients, dose-limiting toxicities, assumed true $p^T$). Cells marked † were not tested and their $p^T$ is interpolated under a monotonicity assumption.

| Adavosertib tier | Olaparib 100 mg | Olaparib 200 mg | Olaparib 300 mg |
|---|---|---|---|
| Low (125–150 mg bid) | 6, 0, 0.00 | 7, 0, 0.00 | 0, 0, 0.10† |
| Medium (175 mg bid) | 4, 0, 0.00 | 35, 3, 0.086 | 5, 1, 0.20 |
| High (200–300 mg qd) | 0, 0, 0.08† | 52, 8, 0.154 | 11, 2, 0.182 |

**Table 9** Case-study replay results (2000 simulated trials of $N = 36$ on the grid of Table 8).

| Design | PCS (%) | NoneSelected (%) |
|---|---|---|
| Ji3+3-Comb | 5.3 | 42.3 |
| Comb-BOIN12 | 11.8 | 0.0 |
| EffTox-approx | 0.1 | 0.1 |

**Regulatory context.** Regulatory acceptance of Bayesian, utility-integrated dose-finding methodology in early-phase oncology combination trials has advanced substantially since the FDA's Project Optimus initiative and its associated dose-optimization guidance formally endorsed moving beyond the maximum-tolerated-dose paradigm (U.S. Food and Drug Administration 2024; Shah et al. 2021; Zirkelbach et al. 2022). Model-assisted Bayesian designs in the BOIN family, methodologically the closest relatives of Comb-BOIN12 and BHUC, already have an established regulatory track record: BOIN-family designs have been used in numerous investigational new drug applications and are valued by reviewers for their transparency, pre-tabulable decision rules, and ease of independent verification (Liu and Yuan 2015; Lin et al. 2020), properties that Ji3+3-Comb and the model-assisted branch of BHUC share by construction. The FDA's more recent draft guidance on the use of Bayesian methodology in clinical trials of drug and biological products (U.S. Food and Drug Administration 2026) explicitly encourages pre-specified Bayesian decision rules of exactly the kind developed here, provided that their operating characteristics, including the behaviour of any information-borrowing mechanism, are documented under a range of concordant and discordant scenarios—precisely the kind of evidence Sections 6–7 of this article provide for BHUC. Because BHUC's cross-indication borrowing is a form of dynamic information sharing across sub-populations, its use in a regulatory submission would also need to satisfy the same type-I-error and dilution-of-evidence considerations that apply to master protocols and platform trials borrowing across cohorts (Schmidli et al. 2014); the explicit, exact finite-sample ceiling on borrowed influence derived in Proposition 4.2, together with the honest disclosure of the mechanism's incomplete discounting at typical early-phase sample sizes (Proposition 4.1 and the remark following it), is intended to give reviewers exactly the kind of pre-specified, quantifiable worst-case bound that regulatory review of a borrowing mechanism requires (International Council for Harmonisation of Technical Requirements for Pharmaceuticals for Human Use (ICH) 2026; Mukherjee et al. 2025). We therefore expect model-assisted designs of the kind proposed here to face a comparatively smooth path through regulatory review relative to fully model-based alternatives, whose joint regression surfaces are harder for reviewers to audit independently and whose fragility at realistic early-phase sample sizes we document directly in Section 8.

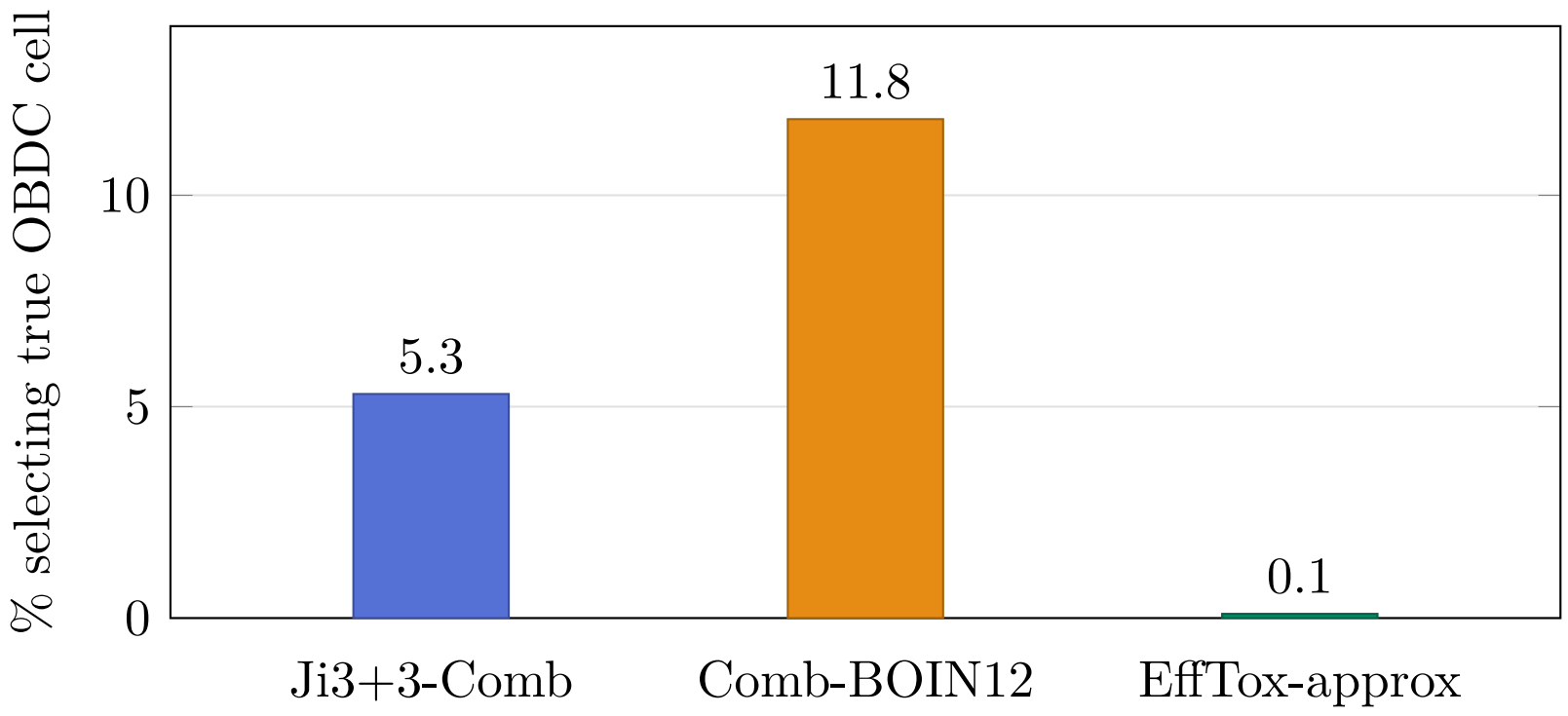

**Figure 6** Correct-selection percentage in the case-study replay of Table 9.

For pharmaceutical and trial biostatisticians planning a two-agent early-phase oncology combination trial, our recommendations, summarized in Figure 2, follow directly from the simulation, sensitivity, and case-study results and are intended to be directly actionable at the protocol-design stage. Model-assisted designs (Comb-BOIN12, and BHUC when multiple indications are involved) offered the most consistently favourable safety–efficacy trade-off across our six single-indication scenarios and in the real-data-grounded case study, and we recommend them as the default choice for two-agent OBDC trials at typical early-phase sample sizes. Rule-based Ji3+3-Comb is preferable when full protocol-level transparency is paramount, provided the trial team accepts a materially higher chance of declaring no acceptable dose when the true efficacy surface plateaus rather than increasing monotonically. Model-based designs remain valuable when the dose grid is large, prior information about the dose–response shape is strong, and dedicated statistical support can monitor the fitted model in real time, but our case study shows concretely that they should not be adopted by default at the modest sample sizes typical of single-indication early-phase trials.

Several limitations, and corresponding future directions, follow. As noted in Section 3, our simulation study generates each patient's toxicity and efficacy outcomes as conditionally independent given dose; real combination regimens can induce within-patient association between the two outcomes (for example when a single shared pharmacological mechanism drives both), and modelling this explicitly, for instance through a bivariate copula linking the toxicity and efficacy margins (Yin and Yuan 2009), would let future work quantify how much within-patient association affects the operating characteristics reported here. As discussed in Section 4.3 (Proposition 4.2 and the remark following it), BHUC's discordance-driven down-weighting is a large-sample property that is only partially realized at the modest per-indication sample sizes typical of early-phase trials, and is entirely absent at a cell's first cohort, where the design necessarily relies on an equally weighted blend of own and borrowed information; trial teams should budget for this by treating the borrowing as a partial rather than complete safeguard against indication discordance in the early part of a trial. Our simulation scenarios and case study use a common $4 \times 4$ or $3 \times 3$ grid and binary toxicity/efficacy endpoints; extending the taxonomy and the BHUC borrowing mechanism to ordinal or time-to-event endpoints, in the spirit of the quasi-binomial six-category construction underlying COMIC (Chen et al. 2025), and to continuous dose curves (Tighiouart et al. 2017; Jimenez and Tighiouart 2022), is a natural next step. BHUC's robust two-component mixture is well suited to the two-or-few-indication regime typical of early combination development; as a regimen matures across a larger number of indications, or as historical data from completed indications accumulate, a fully exchangeable hierarchical random-effects model with an estimated between-indication variance component, fitted by Markov chain Monte Carlo, becomes both identifiable and preferable, and formalizing the transition between these two hierarchical constructions as evidence accrues is an important open problem. Our model-based benchmark uses a Laplace approximation for computational tractability at the scale of a 5000-replication simulation study; incorporating full posterior inference, patient-level covariates through Bayesian additive regression trees (Zhao et al. 2024; Chung et al. 2025; Chen et al. 2024), or pharmacokinetic/pharmacodynamic covariates directly into the utility function would align the model-based branch of the taxonomy more closely with the individualized dosing goals of Project Optimus. Beyond incorporating such covariates, artificial intelligence and machine learning may also enhance how efficiently a trial identifies the OBDC in the first place: reinforcement-learning-based dose-allocation policies, Gaussian-process or Bayesian-optimization surrogates for the joint toxicity–efficacy surface, and meta-learning or transfer-learning approaches that construct informative, cross-trial priors from historical dose-finding databases are all promising directions for reducing the sample size needed to reliably identify the OBDC relative to the grid-search logic underlying all four designs compared here, particularly as dose grids grow larger or combination regimens involve more than two agents; realizing these gains in a regulatory setting will itself require the same kind of transparent, quantifiable operating-characteristic documentation we provide for BHUC in this article, so that an AI/ML-enhanced design remains auditable rather than an opaque black box. Finally, prospective validation of Ji3+3-Comb and BHUC in an operating simulated trial submitted to regulators, following the estimands framework of ICH E9(R1) (International Council for Harmonisation of Technical Requirements for Pharmaceuticals for Human Use (ICH) 2019; Mukherjee et al. 2025) and the model-informed drug development principles of ICH M15 (International Council for Harmonisation of Technical Requirements for Pharmaceuticals for Human Use (ICH) 2026), would be a valuable confirmatory step before either design is adopted in a live combination trial.

## Author Contributions

Ayon Mukherjee (AM), Kentaro Takeda (KT) and James M.S. Wason (JMSW) initiated the project. The concept of this research originated with AM and was further developed by the other authors. The first draft of the manuscript was written jointly by all authors and refined by a collaborative effort of AM and JMSW. AM conducted the simulations, which were validated by KT and JMSW. All authors equally contributed to the review and approved the final manuscript.

## Acknowledgements

This work is a result of an industry and academia collaboration between Newcastle University and Astellas Pharma. The authors thank Newcastle University and Astellas Pharma for providing the resources to complete this work. The authors also thank the National Institute for Health and Care Research for providing the funding to conduct this research.

## Funding

This research was supported by Dr. Ayon Mukherjee's and Prof. James M.S. Wason's National Institute for Health and Care Research grant (grant number: NIHR301614).

## Data Availability Statement

No new patient-level data were generated or analysed during this study. The case study of Section 8 uses only aggregated, publicly reported dose-escalation results from Hamilton et al. (2024). The simulation study, sensitivity analyses, and case-study replay are fully reproducible from the R code provided in the Supplementary Material that follows this article (pages 28–47; Sections S1–S7).

## ORCID

Ayon Mukherjee https://orcid.org/0000-0001-5461-1737
Kentaro Takeda https://orcid.org/0000-0002-8584-0521
James M.S. Wason https://orcid.org/0000-0002-4691-126X

## Supplementary Material

Complete, commented R code implementing all four designs (Ji3+3-Comb, Comb-BOIN12, BHUC, EffTox-approx), the six single-indication scenarios, the two cross-indication borrowing pairs, the sensitivity analyses, and the case-study replay reported in Sections 6–8 is provided in the Supplementary Material, which follows this article as a continuation of the same document (pages 28–47; Sections S1–S7, with Tables and Listings numbered S1, S2, ...).

# A Abbreviations

**Table 10** Abbreviations used in this article.

| Abbreviation | Full form |
|---|---|
| BHUC | Bayesian Hierarchical Utility-based Cross-indication (design) |
| BOIN | Bayesian Optimal Interval |
| CAR-T | Chimeric Antigen Receptor T-cell (therapy) |
| COMIC | Bayesian dose optimization design for drug Combination in Multiple Indications with application to CAR-T therapies |
| CRM | Continual Reassessment Method |
| DLT | Dose-Limiting Toxicity |
| FDA | U.S. Food and Drug Administration |
| ICH | International Council for Harmonisation of Technical Requirements for Pharmaceuticals for Human Use |
| Ji3+3 | Joint i3+3 (design) |
| MCMC | Markov Chain Monte Carlo |
| MTD | Maximum Tolerated Dose |
| MTDC | Maximum Tolerated Dose Combination |
| NCT | National Clinical Trial (ClinicalTrials.gov identifier) |
| OBD | Optimal Biological Dose |
| OBDC | Optimal Biological Dose Combination |
| ORR | Objective Response Rate |
| PARP | Poly (ADP-ribose) Polymerase |
| PCS | Percentage of Correct Selection |
| PFS | Progression-Free Survival |
| PK/PD | Pharmacokinetic/Pharmacodynamic |
| RP2D | Recommended Phase 2 Dose |
| uTPI-Comb | Utility-based Toxicity Probability Interval design for Combinations |

# Supplementary Material

## Optimal Biological Dose Combination Finding: A Design Roadmap and Robust Cross-Indication Bayesian Borrowing

Ayon Mukherjee, Kentaro Takeda and James M.S. Wason

This supplement gives the complete, commented R code used to implement all four designs (Ji3+3-Comb, Comb-BOIN12, BHUC, EffTox-approx), generate the six single-indication scenarios and two cross-indication borrowing pairs, run the 5000-replication simulation study, carry out the prior- and model-assumption sensitivity analyses, replay the illustrative case study, benchmark BHUC's robust mixture against a fixed (non-robust) power-prior ablation, and compute the equal-allocation correct-selection ceiling of Section 6.3 of the main text, all as reported in the main text. All code is provided exactly as executed to produce the results in the main article's Tables 2–10 and Figures 1–6, requiring only base R (version ≥4.0) and the MASS package.

### Contents

### S1 Note on closely-spaced admissible-cell utilities under Scenarios S1 and S2

Section 6.2 of the main text notes that correct-selection percentages are modest under Scenarios S1 (monotone) and S2 (plateau) for all three single-indication designs because several admissible cells attain closely spaced utilities under the true dose–response surface, so that distinguishing the single best cell from its near-neighbours requires more information than $N = 36$ patients spread over a $4 \times 4$ grid can reliably provide. Table S1 lists every admissible cell and its true utility under each scenario, computed directly from the probability matrices in `obdc_funcs.R` (Section S3 below).

Under S1, five of the seven admissible cells lie within a utility band of width 0.075 (0.160–0.235). Under S2, the true OBDC at $(3, 3)$ has utility 0.320, only 0.005 above the second-best admissible cell $(3, 2)$ at 0.315—a gap that is, for all practical purposes, statistically indistinguishable at early-phase sample sizes. This is why we report the percentage of patients allocated to the true OBDC (column PatOBDC in Table 3 of the main text) as a complementary, less noise-sensitive operating characteristic alongside percentage of correct selection.

**Table S1** Admissible cells and true utilities under Scenarios S1 and S2. Cell $(a, b)$ denotes agent-1 dose level $a$, agent-2 dose level $b$. The true OBDC (largest $u$) is marked with $^*$.

| Scenario | Cell $(a, b)$ | $p^T_{ab}$ | $p^E_{ab}$ | $u(a, b)$ |
|---|---|---|---|---|
| S1 | $(4, 1)$ | 0.16 | 0.24 | 0.160 |
| | $(3, 2)$ | 0.18 | 0.28 | 0.190 |
| | $(4, 2)$ | 0.27 | 0.35 | 0.215 |
| | $(2, 3)$ | 0.20 | 0.26 | 0.160 |
| | $(3, 3)^*$ | **0.29** | **0.38** | **0.235** |
| | $(1, 4)$ | 0.22 | 0.20 | 0.090 |
| | $(2, 4)$ | 0.31 | 0.33 | 0.175 |
| S2 | $(2, 1)$ | 0.08 | 0.20 | 0.160 |
| | $(3, 1)$ | 0.13 | 0.28 | 0.215 |
| | $(4, 1)$ | 0.19 | 0.30 | 0.205 |
| | $(2, 2)$ | 0.15 | 0.32 | 0.245 |
| | $(3, 2)$ | 0.21 | 0.42 | 0.315 |
| | $(4, 2)$ | 0.28 | 0.44 | 0.300 |
| | $(1, 3)$ | 0.16 | 0.24 | 0.160 |
| | $(2, 3)$ | 0.23 | 0.40 | 0.285 |
| | $(3, 3)^*$ | **0.30** | **0.47** | **0.320** |
| | $(1, 4)$ | 0.24 | 0.26 | 0.140 |
| | $(2, 4)$ | 0.33 | 0.41 | 0.245 |

## S2 Vectorized posterior evaluation for EffTox-approx

The EffTox-approx design (Section 4.4 of the main text) requires, at every interim decision, posterior draws from the Laplace-approximated logistic regressions for toxicity and efficacy to be evaluated across all $J \times K$ cells of the dose grid. A naive implementation loops over each of the $J \times K$ cells individually for each of `ndraw` posterior draws, which is the dominant computational cost of the design; profiling showed this naive loop to be roughly an order of magnitude slower per replication than the other three designs, which would have made a 5000-replication, six-scenario simulation study computationally prohibitive. We instead precompute a design matrix `Xgrid` spanning the whole dose grid once per scenario, and evaluate all cells simultaneously via a single matrix multiplication per posterior draw set, in the helper function `eval_grid()` below (extracted from `obdc_funcs.R`, full listing in Section S3):

**Listing S1.** Vectorized whole-grid posterior evaluation (`eval_grid`), called at every interim decision of `run_efftox_approx`.

```
## design matrix over the full J*K dose-combination grid (column-major,
## matching R's default matrix(.,J,K) layout) -- used to vectorize
## posterior evaluation across all cells at once instead of looping
## cell-by-cell
Xgrid <- cbind(1, rep(astd,times=K), rep(bstd,each=J),
               rep(astd,times=K)*rep(bstd,each=J))

## vectorized whole-grid posterior evaluation: given fitted Laplace
## objects, returns list(admiss=J*K logical matrix, U=J*K utility
## matrix), evaluating all J*K cells simultaneously via one matrix
## multiplication per margin instead of a per-cell loop
eval_grid <- function(fT, fE, ndraw){
  zT  <- MASS::mvrnorm(ndraw, fT$beta, fT$Sigma)
  zE0 <- MASS::mvrnorm(ndraw, fE$beta, fE$Sigma)
  etaT <- zT %*% t(Xgrid); etaE <- zE0 %*% t(Xgrid)   # ndraw x (J*K)
  pT_draw <- 1/(1+exp(-etaT)); pE_draw <- 1/(1+exp(-etaE))
  admiss_vec <- (colMeans(pT_draw>phiT)<=CT) & (colMeans(pE_draw<phiE)<=CE)
  U_vec <- colMeans(pE_draw - wT*pT_draw)
```

```
  list(admiss=matrix(admiss_vec,J,K), U=matrix(U_vec,J,K))
}
```

This reduces the per-replication cost of EffTox-approx by approximately a factor of two relative to a cell-by-cell loop with the same number of posterior draws (`ndraw=300`), bringing the full 5000-replication, six-scenario single-indication simulation study to under six minutes of wall-clock time on a single core, without altering the statistical logic of the design in any way: `eval_grid()` computes exactly the same posterior means and admissibility indicators as the cell-by-cell formulation, only faster.

## S3 Core design functions and scenario definitions (`obdc_funcs.R`)

**Listing S2.** Global parameters, scenario and indication-pair definitions, the four design functions (Ji3+3-Comb, Comb-BOIN12, BHUC, EffTox-approx), their shared utility helpers, and the fixed (non-robust) power-prior ablation of BHUC used in Section S7.

```
##########################################################################
## obdc_sim.R
## Simulation study comparing four OBDC design paradigms:
##   (1) Ji3+3-Comb   -- proposed rule-based (model-free) design
##   (2) Comb-BOIN12  -- existing model-assisted design (Lu et al., 2025)
##   (3) BHUC         -- proposed hierarchical Bayesian model-assisted
##                       design with cross-indication borrowing
##   (4) EffTox-approx-- fast Laplace-approximation implementation of a
##                       model-based joint efficacy-toxicity design
##
## 5000 replications per scenario, as specified in the manuscript.
##########################################################################

set.seed(20260917)

## ----------------------------------------------------------------
## 0. Global design parameters
## ----------------------------------------------------------------
phiT   <- 0.35     # upper toxicity limit
phiE   <- 0.20     # lower efficacy limit
wT     <- 0.5      # toxicity penalty weight in utility U = pE - wT*pT
CT     <- 0.90      # safety admissibility cutoff:  Pr(pT>phiT|D) <= CT
CE     <- 0.90      # efficacy admissibility cutoff: Pr(pE<phiE|D) <= CE
cohort <- 3
Nmax1  <- 36        # max N, single-indication scenarios
Nmax2  <- 30        # max N per indication, two-indication (borrowing) scenarios
Nstar  <- 6         # BOIN sample-size cutoff parameter
eps_ci <- 0.05      # Ji3+3-Comb equivalence margins used for BOTH tox & eff intervals
gammaDiscount <- 0.5  # BHUC borrowing discount factor
nrep   <- 5000

## BOIN escalation/de-escalation boundaries (Liu & Yuan, 2015 closed form)
boin_boundaries <- function(phiT, phi1 = 0.6*phiT, phi2 = 1.4*phiT){
  lam_e <- log((1-phi1)/(1-phiT)) / log((phiT*(1-phi1))/(phi1*(1-phiT)))
  lam_d <- log((1-phiT)/(1-phi2)) / log((phi2*(1-phiT))/(phiT*(1-phi2)))
  c(lam_e = lam_e, lam_d = lam_d)
}
bb <- boin_boundaries(phiT)
lam_e <- bb["lam_e"]; lam_d <- bb["lam_d"]

## ----------------------------------------------------------------
## 1. Dose grid + scenario library (4 x 4 grid; rows = Agent A dose
##    level 1..4 low->high, cols = Agent B dose level 1..4 low->high)
## ----------------------------------------------------------------
J <- 4; K <- 4
```

```
make_scn <- function(pT, pE, name){
  list(pT = pT, pE = pE, name = name,
       U  = pE - wT*pT)
}

## Scenario 1: monotone efficacy & toxicity in both agents (targeted-agent
## combination, informed by MATCHPOINT-trial-scale thresholds; Brock et al., 2017)
pT1 <- matrix(c(0.03,0.07,0.13,0.22,
                0.06,0.12,0.20,0.31,
                0.10,0.18,0.29,0.41,
                0.16,0.27,0.39,0.52), 4,4, byrow=TRUE)
pE1 <- matrix(c(0.05,0.10,0.15,0.20,
                0.10,0.18,0.26,0.33,
                0.18,0.28,0.38,0.44,
                0.24,0.35,0.45,0.50), 4,4, byrow=TRUE)

## Scenario 2: efficacy plateau at moderate exposure (immune-checkpoint-inhibitor
## combination, ranges informed by CheckMate 012; Hellmann et al., 2017 Lancet Oncol)
pT2 <- matrix(c(0.04,0.09,0.16,0.24,
                0.08,0.15,0.23,0.33,
                0.13,0.21,0.30,0.40,
                0.19,0.28,0.37,0.47), 4,4, byrow=TRUE)
pE2 <- matrix(c(0.10,0.18,0.24,0.26,
                0.20,0.32,0.40,0.41,
                0.28,0.42,0.47,0.47,
                0.30,0.44,0.47,0.46), 4,4, byrow=TRUE)

## Scenario 3: interaction "ridge/valley" -- efficacy dips at the interior
## of the grid due to antagonism at intermediate dose ratios
pT3 <- pT1
pE3 <- matrix(c(0.06,0.12,0.14,0.16,
                0.12,0.16,0.14,0.24,
                0.16,0.18,0.22,0.34,
                0.20,0.24,0.30,0.44), 4,4, byrow=TRUE)

## Scenario 4: high background toxicity -- only the low-dose corner is safe
pT4 <- matrix(c(0.10,0.20,0.33,0.46,
                0.18,0.30,0.44,0.58,
                0.28,0.42,0.55,0.66,
                0.38,0.52,0.63,0.72), 4,4, byrow=TRUE)
pE4 <- matrix(c(0.12,0.22,0.30,0.34,
                0.20,0.32,0.40,0.42,
                0.26,0.38,0.44,0.45,
                0.30,0.40,0.44,0.44), 4,4, byrow=TRUE)

## Scenario 5: no acceptable dose combination exists (all cells overly toxic:
## every cell exceeds phiT=0.35, so a design should ideally recommend no dose)
pT5 <- matrix(pmin(0.99, pT1*1.3+0.38), 4,4)
pE5 <- pE1

## Scenario 6: low toxicity throughout, efficacy monotone -- OBDC at the
## highest dose pair (MTDC coincides with the OBDC)
pT6 <- matrix(c(0.02,0.04,0.06,0.09,
                0.03,0.06,0.09,0.13,
                0.05,0.09,0.13,0.18,
                0.07,0.12,0.17,0.24), 4,4, byrow=TRUE)
pE6 <- matrix(c(0.08,0.14,0.20,0.26,
                0.14,0.22,0.30,0.37,
                0.20,0.30,0.39,0.46,
                0.26,0.37,0.46,0.55), 4,4, byrow=TRUE)

scn_single <- list(
  make_scn(pT1,pE1,"S1: monotone"),
  make_scn(pT2,pE2,"S2: plateau"),
  make_scn(pT3,pE3,"S3: ridge/valley"),
```

```r
  make_scn(pT4,pE4,"S4: high toxicity background"),
  make_scn(pT5,pE5,"S5: all doses overly toxic"),
  make_scn(pT6,pE6,"S6: MTDC = OBDC")
)

## Two-indication scenario PAIRS for the cross-indication borrowing study
## Pair A: concordant indications (similar surfaces -> borrowing should help)
pairA <- list(ind1 = make_scn(pT1, pE1, "Ind.1"),
              ind2 = make_scn(pT1, matrix(pmin(0.95,pE1*1.05+0.02),4,4),
                  "Ind.2 (concordant)"))
## Pair B: discordant indications (different OBDC location -> borrowing should
## be automatically suppressed by the robust mixture)
pE2b <- matrix(c(0.30,0.44,0.47,0.46,
                 0.28,0.42,0.47,0.47,
                 0.20,0.32,0.40,0.41,
                 0.10,0.18,0.24,0.26), 4,4,
                     byrow=TRUE) # OBDC shifted to opposite corner
pairB <- list(ind1 = make_scn(pT1, pE1, "Ind.1"),
              ind2 = make_scn(pT2, pE2b, "Ind.2 (discordant)"))

## ---------------------------------------------------------------
## 2. Utility helpers
## ---------------------------------------------------------------
neighbors <- function(j,k,J,K){
  cand <- rbind(c(j-1,k), c(j+1,k), c(j,k-1), c(j,k+1), c(j,k))
  cand[cand[,1]>=1 & cand[,1]<=J & cand[,2]>=1 & cand[,2]<=K, , drop=FALSE]
}

true_obdc <- function(scn, admiss_only = TRUE){
  ok <- scn$pT <= phiT & scn$pE >= phiE
  if (!any(ok)) return(NULL)
  Um <- scn$U; Um[!ok] <- -Inf
  which(Um==max(Um), arr.ind = TRUE)[1,]
}

## ---------------------------------------------------------------
## 3. Design 1: Ji3+3-Comb  (rule-based / model-free)
## ---------------------------------------------------------------
run_ji3comb <- function(scn, Nmax){
  n  <- matrix(0,J,K); yT <- matrix(0,J,K); yE <- matrix(0,J,K)
  j <- 1; k <- 1; tot <- 0
  visited <- matrix(FALSE,J,K)
  repeat{
    cs <- min(cohort, Nmax-tot)
    if (cs<=0) break
    dT <- rbinom(cs,1,scn$pT[j,k]); dE <- rbinom(cs,1,scn$pE[j,k])
    n[j,k] <- n[j,k]+cs; yT[j,k] <- yT[j,k]+sum(dT); yE[j,k] <- yE[j,k]+sum(dE)
    visited[j,k] <- TRUE
    tot <- tot + cs
    pT_hat <- yT[j,k]/n[j,k]; pE_hat <- yE[j,k]/n[j,k]
    tox_zone <- if (pT_hat > phiT+eps_ci) "O" else if (pT_hat < phiT-eps_ci) "U" else "P"
    eff_zone <- if (pE_hat < phiE-eps_ci) "Low" else "Adequate"
    if (tot>=Nmax) break
    if (tox_zone=="O"){
      opts <- rbind(c(j-1,k), c(j,k-1))
      opts <- opts[opts[,1]>=1 & opts[,2]>=1, , drop=FALSE]
      if (nrow(opts)==0) break  # cannot de-escalate further -> stop (unsafe grid)
      rates <- apply(opts,1,function(o) if(n[o[1],o[2]]>0) yT[o[1],o[2]]/n[o[1],o[2]]
↪ else -1)
      pick <- opts[which.min(rates),]
      j <- pick[1]; k <- pick[2]
    } else if (eff_zone=="Low"){
      opts <- rbind(c(j+1,k), c(j,k+1))
      opts <- opts[opts[,1]<=J & opts[,2]<=K, , drop=FALSE]
      if (nrow(opts)==0){ next } # already at max dose, stay
```

```r
      rates <- apply(opts,1,function(o) if(n[o[1],o[2]]>0) yT[o[1],o[2]]/n[o[1],o[2]]
↪ else 0)
      pick <- opts[which.min(rates),]
      j <- pick[1]; k <- pick[2]
    } # else: stay at (j,k)
  }
  ## rule-based terminal selection: highest observed utility among cells
  ## with acceptable observed toxicity & adequate observed efficacy
  cand <- which(visited & n>0 & (yT/pmax(n,1))<=phiT+eps_ci & (yE/pmax(n,1))>=phiE-eps_ci
↪ ,
      arr.ind=TRUE)
  if (nrow(cand)==0) return(list(sel=c(NA,NA), n=n))
  uhat <- apply(cand,1,function(o) yE[o[1],o[2]]/n[o[1],o[2]] - wT*yT[o[1],o[2]]/n[o[1],o
↪ [2]])
  sel <- cand[which.max(uhat),]
  list(sel=sel, n=n)
}

## ---------------------------------------------------------------
## 4. Design 2: Comb-BOIN12 (model-assisted; Lu, Zhang, Yuan & Lin, 2025)
## ---------------------------------------------------------------
run_comboin12 <- function(scn, Nmax, prior_alpha=NULL, prior_beta=NULL){
  n  <- matrix(0,J,K); yT <- matrix(0,J,K); yE <- matrix(0,J,K)
  j <- 1; k <- 1; tot <- 0
  repeat{
    cs <- min(cohort, Nmax-tot)
    if (cs<=0) break
    dT <- rbinom(cs,1,scn$pT[j,k]); dE <- rbinom(cs,1,scn$pE[j,k])
    n[j,k] <- n[j,k]+cs; yT[j,k] <- yT[j,k]+sum(dT); yE[j,k] <- yE[j,k]+sum(dE)
    tot <- tot + cs
    if (tot>=Nmax) break
    pT_hat <- yT[j,k]/n[j,k]
    if (pT_hat > lam_d){
      ## toxicity above the de-escalation boundary: de-escalation is
      ## mandatory; the utility statistic only chooses the axis
      opts <- rbind(c(j-1,k), c(j,k-1))
      opts <- opts[opts[,1]>=1 & opts[,2]>=1, , drop=FALSE]
      if (nrow(opts)==0) break
      util <- apply(opts,1,function(o) post_util_prob(o,yE,yT,n))
      pick <- opts[which.max(util),]
    } else if (pT_hat<=lam_e){
      ## toxicity at/below the escalation boundary: escalation is
      ## mandatory (room remains to seek more efficacy); the utility
      ## statistic only chooses which agent's dose to raise
      opts <- rbind(c(j+1,k), c(j,k+1))
      opts <- opts[opts[,1]<=J & opts[,2]<=K, , drop=FALSE]
      if (nrow(opts)==0){ pick <- c(j,k) } else {
        util <- apply(opts,1,function(o) post_util_prob(o,yE,yT,n))
        pick <- opts[which.max(util),]
      }
    } else if (n[j,k]>=Nstar){
      ## within the BOIN "stay" interval with an adequately sampled cell
      ## (Comb-BOIN12 Step 2(b)): refine among {stay, de-escalate} only
      opts <- rbind(c(j-1,k), c(j,k-1), c(j,k))
      opts <- opts[opts[,1]>=1 & opts[,2]>=1, , drop=FALSE]
      util <- apply(opts,1,function(o) post_util_prob(o,yE,yT,n))
      pick <- opts[which.max(util),]
    } else {
      ## within the stay interval but under-sampled: stay to accrue
      ## enough information for a reliable toxicity classification
      pick <- c(j,k)
    }
    j <- pick[1]; k <- pick[2]
  }
  sel <- select_obdc(yE,yT,n)
```

```
  list(sel=sel, n=n)
}

## Utility benchmark u_b: the utility value at the pre-specified
## acceptability boundary (pE=phiE, pT=phiT), i.e. the minimum clinically
## acceptable utility. Candidate cells are ranked by their posterior
## probability of exceeding this pre-specified, scenario-independent
## benchmark (formula (3.3) of Lu, Zhang, Yuan & Lin, 2025), rather than by
## a benchmark calibrated against the unknown, unreachable (pE=1,pT=0)
## corner, which would be uninformative whenever the achievable utility
## surface is well below 1 (e.g. under a plateauing dose-efficacy curve).
u_b_bench   <- phiE - wT*phiT
ustar_b     <- (u_b_bench + wT)/(1+wT)

post_util_mean_terminal <- function(o, yE, yT, n, a=1, b=1){
  ## E[u | D_{j,k}] under the quasi-binomial Beta(a,b) model, restricted to
  ## TESTED cells; used only for the terminal OBDC selection.
  jj<-o[1]; kk<-o[2]
  nn <- n[jj,kk]
  if (nn==0) return(-Inf)
  x  <- (yE[jj,kk] + wT*(nn - yT[jj,kk])) / (1+wT)
  post_mean_ustar <- (a+x)/(a+b+nn)
  post_mean_ustar*(1+wT) - wT
}
post_util_mean <- post_util_mean_terminal  # retained for backward compatibility

post_util_prob <- function(o, yE, yT, n, a=1, b=1){
  ## In-trial candidate-ranking statistic: Pr(u*_{j,k} > u*_b | D_{j,k})
  ## (formula (3.3) of Lu, Zhang, Yuan & Lin, 2025). For an as-yet-untested
  ## neighbouring cell (n=0) this correctly reduces to the fixed PRIOR
  ## probability of exceeding the benchmark, so untested cells remain
  ## eligible for exploration; a currently-favoured cell whose accruing data
  ## indicate it is only mediocre will see this probability fall, so the
  ## design does not become permanently anchored to it.
  jj<-o[1]; kk<-o[2]
  nn <- n[jj,kk]
  x  <- if (nn==0) 0 else (yE[jj,kk] + wT*(nn - yT[jj,kk])) / (1+wT)
  pbeta(ustar_b, a+x, b+nn-x, lower.tail = FALSE)
}

select_obdc <- function(yE, yT, n, a=1, b=1){
  ok <- matrix(FALSE,J,K)
  for (jj in 1:J) for (kk in 1:K){
    if (n[jj,kk]==0) next
    ok[jj,kk] <- pbeta(phiT, a+yT[jj,kk], b+n[jj,kk]-yT[jj,kk],
        lower.tail=FALSE) <= CT &&
                 pbeta(phiE, a+yE[jj,kk], b+n[jj,kk]-yE[jj,kk], lower.tail=TRUE)  <= CE
  }
  if (!any(ok)) return(c(NA,NA))
  util <- matrix(-Inf,J,K)
  idx <- which(ok, arr.ind=TRUE)
  for (r in 1:nrow(idx)) util[idx[r,1],idx[r,2]] <- post_util_mean(idx[r,],yE,yT,n)
  wm <- which(util==max(util),arr.ind=TRUE)[1,]
  wm
}

## ------------------------------------------------------------------
## 5. Design 3: BHUC -- proposed hierarchical Bayesian model-assisted
##    design with robust cross-indication borrowing (two indications)
## ------------------------------------------------------------------
## Beta-Binomial (quasi-likelihood) log marginal likelihood, continuised
log_bb_marglik <- function(x, n, a, b){
  lbeta(a+x, b+n-x) - lbeta(a,b)
}
```

```r
run_bhuc_pair <- function(scnA, scnB, Nmax){
  res <- list()
  for (ind in c("A","B")){
    assign(paste0("n",ind), matrix(0,J,K)); assign(paste0("yT",ind), matrix(0,J,K));
        assign(paste0("yE",ind),
        matrix(0,J,K))
  }
  nA<-matrix(0,J,K); yTA<-matrix(0,J,K); yEA<-matrix(0,J,K)
  nB<-matrix(0,J,K); yTB<-matrix(0,J,K); yEB<-matrix(0,J,K)
  jA<-1; kA<-1; jB<-1; kB<-1; totA<-0; totB<-0
  active <- c(TRUE,TRUE)
  repeat{
    if (totA<Nmax){
      cs <- min(cohort, Nmax-totA)
      dT <- rbinom(cs,1,scnA$pT[jA,kA]); dE <- rbinom(cs,1,scnA$pE[jA,kA])
      nA[jA,kA]<-nA[jA,kA]+cs; yTA[jA,kA]<-yTA[jA,kA]+sum(dT);
          yEA[jA,kA]<-yEA[jA,kA]+sum(dE)
      totA <- totA+cs
    }
    if (totB<Nmax){
      cs <- min(cohort, Nmax-totB)
      dT <- rbinom(cs,1,scnB$pT[jB,kB]); dE <- rbinom(cs,1,scnB$pE[jB,kB])
      nB[jB,kB]<-nB[jB,kB]+cs; yTB[jB,kB]<-yTB[jB,kB]+sum(dT);
          yEB[jB,kB]<-yEB[jB,kB]+sum(dE)
      totB <- totB+cs
    }
    if (totA>=Nmax & totB>=Nmax) break
    ## --- indication A step (same mandatory-direction structure as
    ## Comb-BOIN12; the utility statistic bhuc_util() only chooses the axis
    ## within a direction, or refines within the stay zone) ---
    if (totA<Nmax){
      pT_hat <- yTA[jA,kA]/nA[jA,kA]
      if (pT_hat > lam_d){
        opts <- rbind(c(jA-1,kA), c(jA,kA-1));
            opts <- opts[opts[,1]>=1 & opts[,2]>=1,,drop=FALSE]
        if (nrow(opts)>0){ util<-apply(opts,1,bhuc_util,yE=yEA,yT=yTA,n=nA,yEo=yEB,yTo=
↪ yTB,no=nB);
            pick<-opts[which.max(util),]; jA<-pick[1]; kA<-pick[2] }
      } else if (pT_hat<=lam_e){
        opts <- rbind(c(jA+1,kA), c(jA,kA+1));
            opts <- opts[opts[,1]<=J & opts[,2]<=K,,drop=FALSE]
        if (nrow(opts)>0){ util<-apply(opts,1,bhuc_util,yE=yEA,yT=yTA,n=nA,yEo=yEB,yTo=
↪ yTB,no=nB);
            pick<-opts[which.max(util),]; jA<-pick[1]; kA<-pick[2] }
      } else if (nA[jA,kA]>=Nstar){
        opts <- rbind(c(jA-1,kA),c(jA,kA-1),c(jA,kA));
            opts<-opts[opts[,1]>=1 & opts[,2]>=1,,drop=FALSE]
        util<-apply(opts,1,bhuc_util,yE=yEA,yT=yTA,n=nA,yEo=yEB,yTo=yTB,no=nB);
            pick<-opts[which.max(util),];
            jA<-pick[1]; kA<-pick[2]
      } ## else: stay (under-sampled cell within the stay interval)
    }
    ## --- indication B step ---
    if (totB<Nmax){
      pT_hat <- yTB[jB,kB]/nB[jB,kB]
      if (pT_hat > lam_d){
        opts <- rbind(c(jB-1,kB), c(jB,kB-1));
            opts <- opts[opts[,1]>=1 & opts[,2]>=1,,drop=FALSE]
        if (nrow(opts)>0){ util<-apply(opts,1,bhuc_util,yE=yEB,yT=yTB,n=nB,yEo=yEA,yTo=
↪ yTA,no=nA);
            pick<-opts[which.max(util),]; jB<-pick[1]; kB<-pick[2] }
      } else if (pT_hat<=lam_e){
        opts <- rbind(c(jB+1,kB), c(jB,kB+1));
            opts <- opts[opts[,1]<=J & opts[,2]<=K,,drop=FALSE]
        if (nrow(opts)>0){ util<-apply(opts,1,bhuc_util,yE=yEB,yT=yTB,n=nB,yEo=yEA,yTo=
```

```
↪ yTA,no=nA);
          pick<-opts[which.max(util),]; jB<-pick[1]; kB<-pick[2] }
      } else if (nB[jB,kB]>=Nstar){
        opts <- rbind(c(jB-1,kB),c(jB,kB-1),c(jB,kB));
            opts<-opts[opts[,1]>=1 & opts[,2]>=1,,drop=FALSE]
        util<-apply(opts,1,bhuc_util,yE=yEB,yT=yTB,n=nB,yEo=yEA,yTo=yTA,no=nA);
            pick<-opts[which.max(util),];
            jB<-pick[1]; kB<-pick[2]
      } ## else: stay
    }
  }
  selA <- select_obdc_bhuc(yEA,yTA,nA,yEB,yTB,nB)
  selB <- select_obdc_bhuc(yEB,yTB,nB,yEA,yTA,nA)
  list(selA=selA, selB=selB, nA=nA, nB=nB)
}

## posterior mean utility under the robust two-component mixture prior.
## At an as-yet-untested cell (nn=0), the marginal likelihood of "zero
## observations" is exactly 1 (logml=0) under either component for any
## hyperparameters, so this is not a special case: the general formula below
## correctly returns the un-updated 50/50 mixture of the vague and
## discounted-pooled component means, since there is no own-indication
## evidence yet to favour one component over the other.
bhuc_util <- function(o, yE, yT, n, yEo, yTo, no, a0=1, b0=1){
  jj<-o[1]; kk<-o[2]
  nn <- n[jj,kk]
  x  <- (yE[jj,kk] + wT*(nn - yT[jj,kk])) / (1+wT)
  ## component 1: vague, own-data only
  a1 <- a0 + x; b1 <- b0 + (nn - x)
  logml1 <- log_bb_marglik(x, nn, a0, b0)
  ## component 2: pooled prior informed by other indication (discounted)
  no_n <- no[jj,kk]
  if (no_n>0){
    xo <- (yEo[jj,kk] + wT*(no_n - yTo[jj,kk]))/(1+wT)
    a_pool <- a0 + gammaDiscount*xo; b_pool <- b0 + gammaDiscount*(no_n-xo)
  } else { a_pool <- a0; b_pool <- b0 }
  a2 <- a_pool + x; b2 <- b_pool + (nn-x)
  logml2 <- log_bb_marglik(x, nn, a_pool, b_pool)
  ## equal prior weight (0.5/0.5) on the two components -> posterior weight
  m <- max(logml1, logml2)
  w1 <- exp(logml1-m); w2 <- exp(logml2-m)
  post_w2 <- w2/(w1+w2)
  mean1 <- a1/(a1+b1); mean2 <- a2/(a2+b2)
  post_mean_ustar <- (1-post_w2)*mean1 + post_w2*mean2
  post_mean_ustar*(1+wT) - wT
}

select_obdc_bhuc <- function(yE,yT,n,yEo,yTo,no, a=1,b=1){
  ok <- matrix(FALSE,J,K)
  for (jj in 1:J) for (kk in 1:K){
    if (n[jj,kk]==0) next
    ok[jj,kk] <- pbeta(phiT, a+yT[jj,kk], b+n[jj,kk]-yT[jj,kk],
        lower.tail=FALSE) <= CT &&
               pbeta(phiE, a+yE[jj,kk], b+n[jj,kk]-yE[jj,kk], lower.tail=TRUE)  <= CE
  }
  if (!any(ok)) return(c(NA,NA))
  util <- matrix(-Inf,J,K)
  idx <- which(ok, arr.ind=TRUE)
  for (r in 1:nrow(idx)) util[idx[r,1],idx[r,2]] <- bhuc_util(idx[r,],yE,yT,n,yEo,yTo,no)
  which(util==max(util),arr.ind=TRUE)[1,]
}

## ----------------------------------------------------------------
## 6. Design 4: EffTox-approx -- fast Laplace-approximation
##    implementation of a model-based joint efficacy-toxicity design
```

```r
## ---------------------------------------------------------------
## Standardised dose scores (mean 0, sd 0.5), non-collinear interaction a_j*b_k
astd <- scale(1:J)[,1]*0.5/sd(scale(1:J)[,1]);
    bstd <- scale(1:K)[,1]*0.5/sd(scale(1:K)[,1])
## design matrix over the full J*K dose-combination grid (column-major,
## matching R's default matrix(.,J,K) layout) -- used to vectorize posterior
## evaluation across all cells at once instead of looping cell-by-cell
Xgrid <- cbind(1, rep(astd,times=K), rep(bstd,each=J),
    rep(astd,times=K)*rep(bstd,each=J))

fit_laplace <- function(y, Xmat, ridge=1){
  ## penalised (ridge) logistic regression MAP + Laplace covariance
  p <- ncol(Xmat)
  beta <- rep(0,p)
  for (it in 1:25){
    eta <- Xmat%*%beta
    mu <- 1/(1+exp(-eta))
    W <- as.numeric(mu*(1-mu))
    grad <- t(Xmat)%*%(y-mu) - ridge*beta
    H <- -(t(Xmat)%*%(Xmat*W)) - diag(ridge,p)
    step <- tryCatch(solve(H, grad), error=function(e) rep(0,p))
    beta <- beta - step
    if (max(abs(step))<1e-6) break
  }
  eta <- Xmat%*%beta; mu <- 1/(1+exp(-eta)); W <- as.numeric(mu*(1-mu))
  H <- -(t(Xmat)%*%(Xmat*W)) - diag(ridge,p)
  Sigma <- solve(-H)
  list(beta=beta, Sigma=Sigma)
}

## vectorized whole-grid posterior evaluation: given fitted Laplace objects,
## returns list(admiss=J*K logical matrix, U=J*K utility matrix), evaluating
## all J*K cells simultaneously via one matrix multiplication per margin
## instead of a per-cell loop
eval_grid <- function(fT, fE, ndraw){
  zT  <- MASS::mvrnorm(ndraw, fT$beta, fT$Sigma)
  zE0 <- MASS::mvrnorm(ndraw, fE$beta, fE$Sigma)
  etaT <- zT %*% t(Xgrid); etaE <- zE0 %*% t(Xgrid)   # ndraw x (J*K)
  pT_draw <- 1/(1+exp(-etaT)); pE_draw <- 1/(1+exp(-etaE))
  admiss_vec <- (colMeans(pT_draw>phiT)<=CT) & (colMeans(pE_draw<phiE)<=CE)
  U_vec <- colMeans(pE_draw - wT*pT_draw)
  list(admiss=matrix(admiss_vec,J,K), U=matrix(U_vec,J,K))
}

run_efftox_approx <- function(scn, Nmax, ndraw=300){
  n  <- matrix(0,J,K); yT <- matrix(0,J,K); yE <- matrix(0,J,K)
  j <- 1; k <- 1; tot <- 0
  Alog <- numeric(Nmax); Blog <- numeric(Nmax); ABlog <- numeric(Nmax)
  Tlog <- integer(Nmax); Elog <- integer(Nmax)
  repeat{
    cs <- min(cohort, Nmax-tot)
    if (cs<=0) break
    dT <- rbinom(cs,1,scn$pT[j,k]); dE <- rbinom(cs,1,scn$pE[j,k])
    n[j,k]<-n[j,k]+cs; yT[j,k]<-yT[j,k]+sum(dT); yE[j,k]<-yE[j,k]+sum(dE)
    idx <- (tot+1):(tot+cs)
    Alog[idx]<-astd[j]; Blog[idx]<-bstd[k]; ABlog[idx]<-astd[j]*bstd[k]
    Tlog[idx]<-dT; Elog[idx]<-dE
    tot <- tot+cs
    if (tot>=Nmax) break
    XT <- cbind(1, Alog[1:tot], Blog[1:tot], ABlog[1:tot])
    XE <- XT
    fT <- fit_laplace(Tlog[1:tot], XT); fE <- fit_laplace(Elog[1:tot], XE)
    ## posterior draws (independent normal approx for each margin, evaluated
    ## simultaneously across the whole dose grid via eval_grid())
    g <- eval_grid(fT, fE, ndraw)
```

```
    admiss <- g$admiss; Umat <- g$U
    if (!any(admiss)){
      ## global de-escalation toward (1,1) if nothing admissible
      j <- max(1,j-1); k <- max(1,k-1)
    } else {
      cand <- neighbors(j,k,J,K)
      cand <- cand[apply(cand,1,function(o) admiss[o[1],o[2]]), , drop=FALSE]
      if (nrow(cand)==0){
        idx2 <- which(admiss, arr.ind=TRUE)
        util <- apply(idx2,1,function(o) Umat[o[1],o[2]])
        pick <- idx2[which.max(util),]
      } else {
        util <- apply(cand,1,function(o) Umat[o[1],o[2]])
        pick <- cand[which.max(util),]
      }
      j<-pick[1]; k<-pick[2]
    }
  }
  ## final selection: refit once more on complete data
  XT <- cbind(1, Alog[1:tot], Blog[1:tot], ABlog[1:tot])
  fT <- fit_laplace(Tlog[1:tot], XT); fE <- fit_laplace(Elog[1:tot], XT)
  g <- eval_grid(fT, fE, ndraw)
  Umat <- g$U; admiss <- g$admiss
  Umat[n==0] <- -Inf; admiss[n==0] <- FALSE
  if (!any(admiss)) return(list(sel=c(NA,NA), n=n))
  Umask <- Umat; Umask[!admiss] <- -Inf
  sel <- which(Umask==max(Umask),arr.ind=TRUE)[1,]
  list(sel=sel, n=n)
}
library(MASS)

## ----------------------------------------------------------------
## 7. Non-robust ablation: fixed power-prior cross-indication borrowing
##    (same gamma, same sequential rule, no robust vague component;
##    see Section~S7 for the comparison driver)
## ----------------------------------------------------------------
## Fixed (non-robust) power-prior borrowing: same gamma discount and same
## sequential dose-movement / admissibility rules as BHUC, but the utility
## statistic ALWAYS uses the single discounted-pooled Beta posterior -- no
## robust two-component mixture, no discordance-driven down-weighting.
## This isolates robustification as the sole experimental difference from
## BHUC, holding everything else (gamma, escalation rule, admissibility,
## terminal-selection rule) identical.
powerprior_util <- function(o, yE, yT, n, yEo, yTo, no, a0=1, b0=1){
  jj<-o[1]; kk<-o[2]
  nn <- n[jj,kk]
  x  <- (yE[jj,kk] + wT*(nn - yT[jj,kk])) / (1+wT)
  no_n <- no[jj,kk]
  if (no_n>0){
    xo <- (yEo[jj,kk] + wT*(no_n - yTo[jj,kk]))/(1+wT)
    a_pool <- a0 + gammaDiscount*xo; b_pool <- b0 + gammaDiscount*(no_n-xo)
  } else { a_pool <- a0; b_pool <- b0 }
  a2 <- a_pool + x; b2 <- b_pool + (nn-x)
  post_mean_ustar <- a2/(a2+b2)
  post_mean_ustar*(1+wT) - wT
}

select_obdc_powerprior <- function(yE,yT,n,yEo,yTo,no, a=1,b=1){
  ok <- matrix(FALSE,J,K)
  for (jj in 1:J) for (kk in 1:K){
    if (n[jj,kk]==0) next
    ok[jj,kk] <- pbeta(phiT, a+yT[jj,kk], b+n[jj,kk]-yT[jj,kk],
        lower.tail=FALSE) <= CT &&
                 pbeta(phiE, a+yE[jj,kk], b+n[jj,kk]-yE[jj,kk], lower.tail=TRUE)  <= CE
  }
```

```
  if (!any(ok)) return(c(NA,NA))
  util <- matrix(-Inf,J,K)
  idx <- which(ok, arr.ind=TRUE)
  for (r in 1:nrow(idx))
    util[idx[r,1],idx[r,2]] <- powerprior_util(idx[r,],yE,yT,n,yEo,yTo,no)
  which(util==max(util),arr.ind=TRUE)[1,]
}

run_powerprior_pair <- function(scnA, scnB, Nmax){
  nA<-matrix(0,J,K); yTA<-matrix(0,J,K); yEA<-matrix(0,J,K)
  nB<-matrix(0,J,K); yTB<-matrix(0,J,K); yEB<-matrix(0,J,K)
  jA<-1; kA<-1; jB<-1; kB<-1; totA<-0; totB<-0
  repeat{
    if (totA<Nmax){
      cs <- min(cohort, Nmax-totA)
      dT <- rbinom(cs,1,scnA$pT[jA,kA]); dE <- rbinom(cs,1,scnA$pE[jA,kA])
      nA[jA,kA]<-nA[jA,kA]+cs; yTA[jA,kA]<-yTA[jA,kA]+sum(dT);
          yEA[jA,kA]<-yEA[jA,kA]+sum(dE)
      totA <- totA+cs
    }
    if (totB<Nmax){
      cs <- min(cohort, Nmax-totB)
      dT <- rbinom(cs,1,scnB$pT[jB,kB]); dE <- rbinom(cs,1,scnB$pE[jB,kB])
      nB[jB,kB]<-nB[jB,kB]+cs; yTB[jB,kB]<-yTB[jB,kB]+sum(dT);
          yEB[jB,kB]<-yEB[jB,kB]+sum(dE)
      totB <- totB+cs
    }
    if (totA>=Nmax & totB>=Nmax) break
    if (totA<Nmax){
      pT_hat <- yTA[jA,kA]/nA[jA,kA]
      if (pT_hat > lam_d){
        opts <- rbind(c(jA-1,kA), c(jA,kA-1));
            opts <- opts[opts[,1]>=1 & opts[,2]>=1,,drop=FALSE]
        if (nrow(opts)>0){
          util<-apply(opts,1,powerprior_util,yE=yEA,yT=yTA,n=nA,yEo=yEB,yTo=yTB,no=nB)
          pick<-opts[which.max(util),]; jA<-pick[1]; kA<-pick[2] }
      } else if (pT_hat<=lam_e){
        opts <- rbind(c(jA+1,kA), c(jA,kA+1));
            opts <- opts[opts[,1]<=J & opts[,2]<=K,,drop=FALSE]
        if (nrow(opts)>0){
          util<-apply(opts,1,powerprior_util,yE=yEA,yT=yTA,n=nA,yEo=yEB,yTo=yTB,no=nB)
          pick<-opts[which.max(util),]; jA<-pick[1]; kA<-pick[2] }
      } else if (nA[jA,kA]>=Nstar){
        opts <- rbind(c(jA-1,kA),c(jA,kA-1),c(jA,kA));
            opts<-opts[opts[,1]>=1 & opts[,2]>=1,,drop=FALSE]
        util<-apply(opts,1,powerprior_util,yE=yEA,yT=yTA,n=nA,yEo=yEB,yTo=yTB,no=nB)
        pick<-opts[which.max(util),]; jA<-pick[1]; kA<-pick[2]
      }
    }
    if (totB<Nmax){
      pT_hat <- yTB[jB,kB]/nB[jB,kB]
      if (pT_hat > lam_d){
        opts <- rbind(c(jB-1,kB), c(jB,kB-1));
            opts <- opts[opts[,1]>=1 & opts[,2]>=1,,drop=FALSE]
        if (nrow(opts)>0){
          util<-apply(opts,1,powerprior_util,yE=yEB,yT=yTB,n=nB,yEo=yEA,yTo=yTA,no=nA)
          pick<-opts[which.max(util),]; jB<-pick[1]; kB<-pick[2] }
      } else if (pT_hat<=lam_e){
        opts <- rbind(c(jB+1,kB), c(jB,kB+1));
            opts <- opts[opts[,1]<=J & opts[,2]<=K,,drop=FALSE]
        if (nrow(opts)>0){
          util<-apply(opts,1,powerprior_util,yE=yEB,yT=yTB,n=nB,yEo=yEA,yTo=yTA,no=nA)
          pick<-opts[which.max(util),]; jB<-pick[1]; kB<-pick[2] }
      } else if (nB[jB,kB]>=Nstar){
        opts <- rbind(c(jB-1,kB),c(jB,kB-1),c(jB,kB));
```

```
            opts<-opts[opts[,1]>=1 & opts[,2]>=1,,drop=FALSE]
          util<-apply(opts,1,powerprior_util,yE=yEB,yT=yTB,n=nB,yEo=yEA,yTo=yTA,no=nA)
          pick<-opts[which.max(util),]; jB<-pick[1]; kB<-pick[2]
        }
      }
    }
  selA <- select_obdc_powerprior(yEA,yTA,nA,yEB,yTB,nB)
  selB <- select_obdc_powerprior(yEB,yTB,nB,yEA,yTA,nA)
  list(selA=selA, selB=selB, nA=nA, nB=nB)
}
```

## S4 Simulation drivers: single-indication and cross-indication studies

**Listing S3.** Driver script for the 5000-replication single-indication simulation study (Table 3 and Figure 3 of the main text).

```
source("obdc_funcs.R")
set.seed(20260917)
nrep <- 5000

summarise_single <- function(scn, Nmax, nrep, design){
  tru <- true_obdc(scn)
  pcs <- 0; patOBDC <- 0; overdose_sel <- 0; overdose_pat <- 0; meanN <- 0; nonesel <- 0
  for (r in 1:nrep){
    out <- switch(design,
      "Ji3+3-Comb" = run_ji3comb(scn, Nmax),
      "Comb-BOIN12" = run_combboin12(scn, Nmax),
      "EffTox-approx" = run_efftox_approx(scn, Nmax))
    sel <- out$sel; n <- out$n
    meanN <- meanN + sum(n)
    overdose_pat <- overdose_pat + sum(n[scn$pT>phiT])
    if (any(is.na(sel))){ nonesel <- nonesel+1; next }
    if (!is.null(tru) && all(sel==tru)) pcs <- pcs+1
    if (!is.null(tru)) patOBDC <- patOBDC + n[tru[1],tru[2]]
    if (scn$pT[sel[1],sel[2]] > phiT) overdose_sel <- overdose_sel+1
  }
  data.frame(scenario=scn$name, design=design,
             PCS=100*pcs/nrep, PatOBDC=100*patOBDC/nrep/Nmax,
             OverdoseSel=100*overdose_sel/nrep,
                 OverdosePat=100*overdose_pat/(nrep*Nmax),
             MeanN=meanN/nrep, NoneSelected=100*nonesel/nrep)
}

cat("Running single-indication scenarios, nrep=",nrep,"\n")
t00 <- Sys.time()
results_single <- do.call(rbind, lapply(scn_single, function(scn){
  do.call(rbind, lapply(c("Ji3+3-Comb","Comb-BOIN12","EffTox-approx"), function(d){
    t0<-Sys.time()
    r <- summarise_single(scn, Nmax1, nrep, d)
    cat(sprintf("  %s / %s  [%.1fs]\n", scn$name, d,
        as.numeric(Sys.time()-t0,units="secs")))
    r
  }))
}))
print(results_single, row.names=FALSE)
write.csv(results_single, "results_single.csv", row.names=FALSE)
cat(sprintf("TOTAL TIME: %.1fs\n", as.numeric(Sys.time()-t00,units="secs")))
cat("DONE_SINGLES\n")
```

**Listing S4.** Driver script for the 5000-replication cross-indication borrowing study (Table 5 and Figure 4 of the main text). The core two-method (Comb-BOIN12, BHUC) simulation loop is run first and is byte-for-byte identical to the loop used to generate the originally reported Comb-BOIN12/BHUC numbers; the fixed power-prior ablation (Section S3) is appended as a wholly separate block after this loop completes so that introducing a third comparator does not alter the sequence of pseudo-random draws consumed by, and hence does not perturb, the Comb-BOIN12/BHUC results. A further sensitivity sweep of this ablation over the borrowing discount $\gamma$ appears in Section S7.

```
source("obdc_funcs.R")
set.seed(20260917)
nrep <- 5000

summarise_pair <- function(pair, Nmax, nrep, method){
  truA <- true_obdc(pair$ind1); truB <- true_obdc(pair$ind2)
  pcsA<-0; pcsB<-0; nA_tot<-0; nB_tot<-0
  for (r in 1:nrep){
    if (method=="Comb-BOIN12 (no borrowing)"){
      outA <- run_combboin12(pair$ind1, Nmax);
          outB <- run_combboin12(pair$ind2, Nmax)
      selA <- outA$sel; selB <- outB$sel; nA <- outA$n; nB <- outB$n
    } else {
      out <- run_bhuc_pair(pair$ind1, pair$ind2, Nmax)
      selA <- out$selA; selB <- out$selB; nA <- out$nA; nB <- out$nB
    }
    nA_tot <- nA_tot+sum(nA); nB_tot <- nB_tot+sum(nB)
    if (!any(is.na(selA)) && !is.null(truA) && all(selA==truA)) pcsA <- pcsA+1
    if (!any(is.na(selB)) && !is.null(truB) && all(selB==truB)) pcsB <- pcsB+1
  }
  data.frame(method=method, PCS_Ind1=100*pcsA/nrep, PCS_Ind2=100*pcsB/nrep,
             MeanN_Ind1=nA_tot/nrep, MeanN_Ind2=nB_tot/nrep)
}

cat("Running two-indication borrowing scenarios (Comb-BOIN12, BHUC), nrep=",
    nrep,"\n")
t00 <- Sys.time()
core_methods <- c("Comb-BOIN12 (no borrowing)","BHUC (proposed)")
res_pairA <- do.call(rbind, lapply(core_methods, function(m){
  t0<-Sys.time(); r<-summarise_pair(pairA, Nmax2, nrep, m)
  cat(sprintf("  pairA / %s [%.1fs]\n", m,
      as.numeric(Sys.time()-t0,units="secs")))
  r
}))
res_pairA$pair <- "Concordant indications"
res_pairB <- do.call(rbind, lapply(core_methods, function(m){
  t0<-Sys.time(); r<-summarise_pair(pairB, Nmax2, nrep, m)
  cat(sprintf("  pairB / %s [%.1fs]\n", m,
      as.numeric(Sys.time()-t0,units="secs")))
  r
}))
res_pairB$pair <- "Discordant indications"
results_pairs <- rbind(res_pairA,res_pairB)
print(results_pairs,row.names=FALSE)
write.csv(results_pairs, "results_pairs.csv", row.names=FALSE)
cat(sprintf("TOTAL TIME: %.1fs\n", as.numeric(Sys.time()-t00,units="secs")))
cat("DONE_PAIRS_CORE\n")

## Fixed (non-robust) power-prior ablation (Section~S3), appended AFTER the
## block above so the RNG stream already consumed -- and hence the
## Comb-BOIN12 / BHUC numbers just printed -- is byte-for-byte unaffected
## by introducing this third comparator. See Section~S7 for the further
## gamma-sensitivity sweep of this same ablation.
summarise_pair_pp <- function(pair, Nmax, nrep){
  truA <- true_obdc(pair$ind1); truB <- true_obdc(pair$ind2)
  pcsA<-0; pcsB<-0; nA_tot<-0; nB_tot<-0
```

```
  for (r in 1:nrep){
    out <- run_powerprior_pair(pair$ind1, pair$ind2, Nmax)
    nA_tot <- nA_tot+sum(out$nA); nB_tot <- nB_tot+sum(out$nB)
    if (!any(is.na(out$selA)) && !is.null(truA) && all(out$selA==truA))
        pcsA <- pcsA+1
    if (!any(is.na(out$selB)) && !is.null(truB) && all(out$selB==truB))
        pcsB <- pcsB+1
  }
  data.frame(method="Fixed power prior (non-robust)",
             PCS_Ind1=100*pcsA/nrep, PCS_Ind2=100*pcsB/nrep,
             MeanN_Ind1=nA_tot/nrep, MeanN_Ind2=nB_tot/nrep)
}

cat("Running fixed power-prior ablation, nrep=",nrep,"\n")
t01 <- Sys.time()
res_ppA <- summarise_pair_pp(pairA, Nmax2, nrep)
res_ppA$pair <- "Concordant indications"
res_ppB <- summarise_pair_pp(pairB, Nmax2, nrep)
res_ppB$pair <- "Discordant indications"
results_pairs_pp <- rbind(res_ppA,res_ppB)
print(results_pairs_pp,row.names=FALSE)
write.csv(results_pairs_pp, "results_pairs_powerprior.csv", row.names=FALSE)
cat(sprintf("TOTAL TIME: %.1fs\n", as.numeric(Sys.time()-t01,units="secs")))
cat("DONE_PAIRS_POWERPRIOR\n")
```

## S5 Sensitivity analysis code

**Listing S5.** Sensitivity analyses for the BHUC borrowing discount $\gamma$, the utility weight $w_T$, and the quasi-binomial prior Beta$(a_0, b_0)$ (Tables 6–7 and Figure 5 of the main text).

```
source("obdc_funcs.R")
set.seed(20260918)
nrep_s <- 2000

## helper: recompute wT-dependent globals and scenario utility surfaces
set_wT <- function(w){
  wT <<- w
  u_b_bench <<- phiE - wT*phiT
  ustar_b   <<- (u_b_bench + wT)/(1+wT)
}
rescale_scn <- function(scn, w) { scn$U <- scn$pE - w*scn$pT; scn }

## Comb-BOIN12 variant exposing the quasi-binomial Beta(a,b) prior
run_combboin12_prior <- function(scn, Nmax, a=1, b=1){
  n  <- matrix(0,J,K); yT <- matrix(0,J,K); yE <- matrix(0,J,K)
  j <- 1; k <- 1; tot <- 0
  repeat{
    cs <- min(cohort, Nmax-tot); if (cs<=0) break
    dT <- rbinom(cs,1,scn$pT[j,k]); dE <- rbinom(cs,1,scn$pE[j,k])
    n[j,k]<-n[j,k]+cs; yT[j,k]<-yT[j,k]+sum(dT); yE[j,k]<-yE[j,k]+sum(dE)
    tot <- tot+cs; if (tot>=Nmax) break
    pT_hat <- yT[j,k]/n[j,k]
    if (pT_hat > lam_d){
      opts <- rbind(c(j-1,k), c(j,k-1));
          opts<-opts[opts[,1]>=1 & opts[,2]>=1,,drop=FALSE]
      if (nrow(opts)==0) break
      util <- apply(opts,1,function(o) post_util_prob(o,yE,yT,n,a=a,b=b));
          pick<-opts[which.max(util),]
    } else if (pT_hat<=lam_e){
      opts <- rbind(c(j+1,k), c(j,k+1));
          opts<-opts[opts[,1]<=J & opts[,2]<=K,,drop=FALSE]
      if (nrow(opts)==0) pick<-c(j,k) else {
          util<-apply(opts,1,function(o) post_util_prob(o,yE,yT,n,a=a,b=b));
          pick<-opts[which.max(util),] }
```

```
    } else if (n[j,k]>=Nstar){
      opts <- rbind(c(j-1,k), c(j,k-1), c(j,k));
          opts<-opts[opts[,1]>=1 & opts[,2]>=1,,drop=FALSE]
      util <- apply(opts,1,function(o) post_util_prob(o,yE,yT,n,a=a,b=b));
          pick<-opts[which.max(util),]
    } else pick <- c(j,k)
    j<-pick[1]; k<-pick[2]
  }
  sel <- select_obdc(yE,yT,n,a=a,b=b)
  list(sel=sel, n=n)
}

summarise_single_g <- function(scn, Nmax, nrep, runner, wgt){
  tru <- true_obdc(rescale_scn(scn, wgt))
  pcs <- 0; nonesel <- 0; overdose_sel <- 0
  for (r in 1:nrep){
    out <- runner(scn, Nmax); sel <- out$sel
    if (any(is.na(sel))){ nonesel <- nonesel+1; next }
    if (!is.null(tru) && all(sel==tru)) pcs <- pcs+1
    if (scn$pT[sel[1],sel[2]] > phiT) overdose_sel <- overdose_sel+1
  }
  data.frame(PCS=100*pcs/nrep, OverdoseSel=100*overdose_sel/nrep,
      NoneSelected=100*nonesel/nrep)
}

## ---- Axis A: BHUC borrowing-discount (gamma) sensitivity ----
cat("Axis A: BHUC discount gamma sensitivity\n")
resA <- list()
for (g in c(0.25,0.5,0.75)){
  gammaDiscount <<- g
  for (pn in c("pairA","pairB")){
    pr <- get(pn)
    pcsA<-0; pcsB<-0
    truA <- true_obdc(pr$ind1); truB <- true_obdc(pr$ind2)
    for (r in 1:nrep_s){
      out <- run_bhuc_pair(pr$ind1, pr$ind2, Nmax2)
      if(!any(is.na(out$selA)) && !is.null(truA) && all(out$selA==truA)) pcsA<-pcsA+1
      if(!any(is.na(out$selB)) && !is.null(truB) && all(out$selB==truB)) pcsB<-pcsB+1
    }
    resA[[length(resA)+1]] <- data.frame(gamma=g, pair=pn, PCS_Ind1=100*pcsA/nrep_s,
        PCS_Ind2=100*pcsB/nrep_s)
    cat(sprintf("  gamma=%.2f %s done\n", g, pn))
  }
}
gammaDiscount <<- 0.5
res_gamma <- do.call(rbind, resA)
write.csv(res_gamma, "sens_gamma.csv", row.names=FALSE)
print(res_gamma, row.names=FALSE)

## ---- Axis B: utility-weight (wT) model-assumption sensitivity ----
cat("\nAxis B: utility weight wT sensitivity\n")
resB <- list()
for (w in c(0.3,0.5,0.7)){
  set_wT(w)
  for (sidx in c(1,3)){
    scn <- scn_single[[sidx]]
    for (dname in c("Ji3+3-Comb","Comb-BOIN12","EffTox-approx")){
      runner <- switch(dname,
        "Ji3+3-Comb"=function(s,N) run_ji3comb(s,N),
        "Comb-BOIN12"=function(s,N) run_combboin12(s,N),
        "EffTox-approx"=function(s,N) run_efftox_approx(s,N))
      r <- summarise_single_g(scn, Nmax1, nrep_s, runner, w)
      resB[[length(resB)+1]] <- data.frame(wT=w, scenario=scn$name, design=dname, r)
      cat(sprintf("  wT=%.1f %s / %s done\n", w, scn$name, dname))
    }
```

```
  }
}
set_wT(0.5)
res_wT <- do.call(rbind, resB)
write.csv(res_wT, "sens_wT.csv", row.names=FALSE)
print(res_wT, row.names=FALSE)

## ---- Axis C: quasi-binomial Beta(a,b) prior-informativeness sensitivity
## (Comb-BOIN12) ----
cat("\nAxis C: Beta(a,b) prior sensitivity (Comb-BOIN12)\n")
resC <- list()
for (ab in list(c(0.5,0.5),c(1,1),c(2,2))){
  a<-ab[1]; b<-ab[2]
  for (sidx in c(1,4)){
    scn <- scn_single[[sidx]]
    tru <- true_obdc(scn)
    pcs<-0; nonesel<-0; overdose_sel<-0
    for (r in 1:nrep_s){
      out <- run_combboin12_prior(scn, Nmax1, a=a, b=b); sel<-out$sel
      if (any(is.na(sel))){ nonesel<-nonesel+1; next }
      if (!is.null(tru) && all(sel==tru)) pcs<-pcs+1
      if (scn$pT[sel[1],sel[2]] > phiT) overdose_sel <- overdose_sel+1
    }
    resC[[length(resC)+1]] <- data.frame(prior=sprintf("Beta(%.1f,%.1f)",a,b),
        scenario=scn$name,
                                          PCS=100*pcs/nrep_s,
                                              OverdoseSel=100*overdose_sel/nrep_s,
                                              NoneSelected=100*nonesel/nrep_s)
    cat(sprintf("  Beta(%.1f,%.1f) %s done\n", a,b, scn$name))
  }
}
res_prior <- do.call(rbind, resC)
write.csv(res_prior, "sens_prior.csv", row.names=FALSE)
print(res_prior, row.names=FALSE)

cat("\nDONE_SENSITIVITY\n")
```

## S6 Illustrative case study replay code

**Listing S6.** Replay of all three single-indication designs on the $3 \times 3$ grid constructed from Hamilton et al. (2024) (Table 9 and Figure 6 of the main text).

```
source("obdc_funcs.R")
set.seed(20260919)

## Override global grid dimensions for the 3x3 case-study grid derived from
## the adavosertib+olaparib phase Ib trial (Hamilton et al., 2024)
J <- 3; K <- 3
astd <- scale(1:J)[,1]*0.5/sd(scale(1:J)[,1])
bstd <- scale(1:K)[,1]*0.5/sd(scale(1:K)[,1])
Xgrid <- cbind(1, rep(astd,times=K), rep(bstd,each=J),
    rep(astd,times=K)*rep(bstd,each=J))

## Rows = adavosertib intensity tier (T1<T2<T3); columns = olaparib dose
## level (L1=100mg, L2=200mg, L3=300mg bid), matrix filled by row
pT_case <- matrix(c(0.00, 0.00, 0.10,
                     0.00, 0.086, 0.20,
                     0.08, 0.154, 0.182), nrow=3, byrow=TRUE)
pE_case <- matrix(c(0.05, 0.15, 0.12,
                     0.08, 0.308, 0.20,
                     0.07, 0.091, 0.15), nrow=3, byrow=TRUE)
case_scn <- list(pT=pT_case, pE=pE_case, name="Adavosertib+Olaparib case study",
                  U = pE_case - wT*pT_case)
tru <- true_obdc(case_scn)
```

```
cat("True (illustrative) OBDC cell (row=adavosertib tier, col=olaparib level):", tru,
    "\n")
cat("Admissible cells:\n")
print(which(case_scn$pT<=phiT & case_scn$pE>=phiE, arr.ind=TRUE))

Nmax_case <- 36
nrep_case <- 2000
summarise_case <- function(design){
  pcs<-0; nonesel<-0; sel_tab <- matrix(0,J,K)
  for (r in 1:nrep_case){
    out <- switch(design,
      "Ji3+3-Comb"=run_ji3comb(case_scn, Nmax_case),
      "Comb-BOIN12"=run_combboin12(case_scn, Nmax_case),
      "EffTox-approx"=run_efftox_approx(case_scn, Nmax_case))
    sel <- out$sel
    if (any(is.na(sel))) { nonesel <- nonesel+1; next }
    sel_tab[sel[1],sel[2]] <- sel_tab[sel[1],sel[2]]+1
    if (!is.null(tru) && all(sel==tru)) pcs <- pcs+1
  }
  list(PCS=100*pcs/nrep_case, NoneSelected=100*nonesel/nrep_case,
      sel_tab=100*sel_tab/nrep_case)
}
for (d in c("Ji3+3-Comb","Comb-BOIN12","EffTox-approx")){
  r <- summarise_case(d)
  cat(sprintf("\n%s: PCS=%.1f%%  NoneSelected=%.1f%%\n", d, r$PCS, r$NoneSelected))
  print(round(r$sel_tab,1))
}
```

## S7 Robustness checks for the cross-indication borrowing comparison

This section supports the discussion in Section 6.3 of the main text and the literature-comparison subsection of Section 4. It contains two independent pieces of code. The first (Section S7.1) re-runs the fixed power-prior ablation of Section S4 at a second, larger borrowing discount ($\gamma = 0.9$) to check whether the qualitative finding of Section 4—that a fixed, non-robust power prior performs comparably to, and on these two scenarios occasionally modestly better than, BHUC—is an artifact of the specific discount used elsewhere in the paper ($\gamma = 0.5$); it is not. This driver is entirely self-contained (its own `set.seed()` call and its own sequence of random draws), so its numbers are not expected to reproduce Table 5 of the main text exactly; it is a robustness check, not a re-derivation of that table. The second (Section S7.2) computes the equal-allocation correct-selection ceiling reported in Table 4 of the main text.

### S7.1 Discount-parameter sensitivity of the power-prior ablation

**Listing S7.** Sensitivity of the fixed power-prior ablation (Section S4) to the borrowing discount $\gamma$. The concordant pair is evaluated at $\gamma = 0.5$ (the value used everywhere else in the paper); the discordant pair is evaluated at both $\gamma = 0.5$ and a larger discount $\gamma = 0.9$, since a referee might reasonably ask whether more aggressive fixed borrowing changes the qualitative conclusion under discordance. It does not: even at $\gamma = 0.9$ the fixed power prior remains comparable to BHUC on Indication 1 and only modestly behind it on Indication 2, well short of the qualitative separation a naive "robust beats non-robust on average" narrative would predict. This is precisely why the paper grounds BHUC's justification in the distribution-free worst-case guarantee of Proposition 4.2 rather than in an average-case empirical margin.

```
source("obdc_funcs.R")
set.seed(20260917)
nrep <- 5000

summarise_pair_ablation <- function(pair, Nmax, nrep, g){
  gammaDiscount <<- g
```

```
  truA <- true_obdc(pair$ind1); truB <- true_obdc(pair$ind2)
  pcsA_bhuc<-0; pcsB_bhuc<-0; pcsA_pp<-0; pcsB_pp<-0
  for (r in 1:nrep){
    outB <- run_bhuc_pair(pair$ind1, pair$ind2, Nmax)
    if (!any(is.na(outB$selA)) && !is.null(truA) && all(outB$selA==truA))
      pcsA_bhuc <- pcsA_bhuc+1
    if (!any(is.na(outB$selB)) && !is.null(truB) && all(outB$selB==truB))
      pcsB_bhuc <- pcsB_bhuc+1
    outP <- run_powerprior_pair(pair$ind1, pair$ind2, Nmax)
    if (!any(is.na(outP$selA)) && !is.null(truA) && all(outP$selA==truA))
      pcsA_pp <- pcsA_pp+1
    if (!any(is.na(outP$selB)) && !is.null(truB) && all(outP$selB==truB))
      pcsB_pp <- pcsB_pp+1
  }
  data.frame(gamma=g, BHUC_Ind1=100*pcsA_bhuc/nrep, BHUC_Ind2=100*pcsB_bhuc/nrep,
             PowerPrior_Ind1=100*pcsA_pp/nrep, PowerPrior_Ind2=100*pcsB_pp/nrep)
}

cat("Concordant pair (gamma=0.5):\n")
print(summarise_pair_ablation(pairA, Nmax2, nrep, 0.5))
cat("Discordant pair, gamma sweep:\n")
res_disc <- do.call(rbind, lapply(c(0.5, 0.9), function(g)
  summarise_pair_ablation(pairB, Nmax2, nrep, g)))
print(res_disc, row.names=FALSE)
gammaDiscount <<- 0.5
write.csv(res_disc, "results_ablation_gamma.csv", row.names=FALSE)
cat("DONE_ABLATION\n")
```

### S7.2 Equal-allocation correct-selection ceiling

**Listing S8.** Equal-allocation oracle for the correct-selection ceiling under Scenarios S1–S4 (Table 4 of the main text). The oracle is told the true admissible set in advance—so it cannot mis-classify admissibility, unlike every design evaluated elsewhere in the paper—and splits the fixed total sample size $N_{\max}$ equally across exactly those cells, which is the classical optimal allocation for a pure ranking-and-selection problem with a known candidate set. It then applies the same Beta$(1, 1)$ posterior-mean terminal decision rule used by Comb-BOIN12. As documented in the code comments and in Section 6.3 of the main text, this bound is valid only for nonparametric, per-cell decision procedures evaluated under Scenarios S1–S4: it is not a ceiling for EffTox-approx, whose parametric borrowing across cells can exceed a per-cell bound, and it is not meaningful under Scenario S5 (no admissible cell) or Scenario S6 (13 of 16 cells admissible, where adaptive concentration legitimately outperforms an equal split).

```
source("obdc_funcs.R")
set.seed(20260922)

## Equal-allocation oracle: knows the TRUE admissible set in advance (no
## risk of mis-classifying admissibility) and splits the total budget N
## equally among just those cells -- the classical optimal allocation for
## a pure ranking-and-selection problem -- then applies the SAME Beta(1,1)
## posterior-mean terminal rule used by Comb-BOIN12. This upper-bounds the
## correct-selection probability achievable by any NONPARAMETRIC, per-cell
## procedure with the same total N (it is not a bound for EffTox-approx,
## whose parametric model borrows across cells, and is not meaningful
## under S5, where no cell is admissible, or S6, where 13/16 cells are
## admissible and adaptive concentration legitimately outperforms an
## equal split -- see main text, Section 6.3).
oracle_pcs <- function(scn, Nmax, nrep, a=1, b=1){
  ok <- scn$pT <= phiT & scn$pE >= phiE
  if (!any(ok)) return(c(PCS=NA, n_admiss=0, n_each=NA))
  idx <- which(ok, arr.ind=TRUE)
  Kc  <- nrow(idx)
  n_each <- floor(Nmax / Kc)
```

```
  extra  <- Nmax - n_each*Kc
  n_cell <- rep(n_each, Kc); if (extra>0) n_cell[1:extra] <- n_cell[1:extra] + 1
  Um <- scn$U; Um[!ok] <- -Inf
  tru <- which(Um==max(Um), arr.ind=TRUE)[1,]
  pcs <- 0
  for (r in 1:nrep){
    util <- numeric(Kc)
    for (c in 1:Kc){
      jj<-idx[c,1]; kk<-idx[c,2]; nn<-n_cell[c]
      if (nn==0){ util[c] <- -Inf; next }
      dT <- rbinom(1,nn,scn$pT[jj,kk]); dE <- rbinom(1,nn,scn$pE[jj,kk])
      x <- (dE + wT*(nn-dT))/(1+wT)
      post_mean_ustar <- (a+x)/(a+b+nn)
      util[c] <- post_mean_ustar*(1+wT) - wT
    }
    pick <- idx[which.max(util),]
    if (all(pick==tru)) pcs <- pcs+1
  }
  c(PCS=100*pcs/nrep, n_admiss=Kc, n_each=n_each)
}

nrep <- 5000
scn_names <- c("S1: monotone","S2: plateau","S3: ridge/valley",
               "S4: high toxicity background")
res_oracle <- do.call(rbind, lapply(1:4, function(i){
  r <- oracle_pcs(scn_single[[i]], Nmax1, nrep)
  data.frame(scenario=scn_names[i], PCS_oracle=r["PCS"],
             n_admissible=r["n_admiss"], n_per_cell=r["n_each"])
}))
print(res_oracle, row.names=FALSE)
write.csv(res_oracle, "results_oracle.csv", row.names=FALSE)
cat("DONE_ORACLE\n")
```